\documentclass[graybox]{svmult}

\usepackage{mathptmx}       
\usepackage{helvet}         
\usepackage{courier}        
\usepackage{type1cm}        
\usepackage{makeidx}         
\usepackage{graphicx}        
\usepackage{multicol}        
\usepackage[bottom]{footmisc}
\usepackage{url}

\newcommand{\rref}[1]{(\ref{#1})} 
\def\bra<#1|{\mathinner{\langle\,{#1}\,\vert}} 
\def\ket|#1>{\mathinner{\vert\,{#1}\,\rangle}} 
\def\red|#1|{\mathinner{\!\vert\,{#1}\,\vert\!}}
\def\braket<#1>{\mathinner{\langle\,{#1}\,\rangle}}

\makeindex             

\begin{document}

\title*{Atomic density functions: atomic physics calculations analyzed with methods from quantum chemistry}
\author{Alex Borgoo, Michel Godefroid, Paul Geerlings}
\institute{Alex Borgoo \at Vrije Universiteit Brussel, \email{aborgoo@vub.ac.be}
\and Michel R. Godefroid \at Universit\'e Libre de Bruxelles \email{mrgodef@ulb.ac.be}
\and Paul Geerlings \at Vrije Universiteit Brussel \email{pgeerlin@vub.ac.be}}
%
%

\maketitle

\tableofcontents
\newpage
\abstract{This contribution reviews a selection of findings on atomic density functions and discusses ways for reading chemical information from them. First an expression for the density function for atoms in the multi-configuration Hartree--Fock scheme is established. The spherical harmonic content of the density function and ways to restore the spherical symmetry in a general open-shell case are treated. The evaluation of the density function is illustrated in a few examples. In the second part of the paper, atomic density functions are analyzed using quantum similarity measures. The comparison of atomic density functions is shown to be useful to obtain physical and chemical information. Finally, concepts from information theory are introduced and adopted for the comparison of density functions. In particular, based on the Kullback--Leibler form, a functional is constructed that reveals the periodicity in Mendeleev's table. Finally a quantum similarity measure is constructed, based on the integrand of the Kullback--Leibler expression and the periodicity is regained in a different way. }

\section{Introduction}

Density Functional Theory (DFT) plays a prominent role in present day investigations of the electronic structure of atoms and molecules. Within DFT the electron density function plays a central role as it caries all the information to describe the investigated system. The idea that all physical and chemical information is contained in the density incited the present authors to try and recover some of this information. Although the proof of the Hohenberg--Kohn theorems~\cite{HOHENBERG:1964}, which guarantee the presence of all physical information in the density function, is generally said to be disarmingly simple, it does not provide a method to get to the relevant information. The continued search for improved energy functionals is the most evident example to illustrate the challenge researchers are confronted with.

In recent years, the present authors have developed an interest in obtaining chemical information from atonic density functions. The application of concepts from quantum chemistry shows that some  particular aspects of physical and chemical interest can be read from the density functions. In particular the comparison of density functions using quantum similarity measures or functionals from information theory plays an important role. The original goal of the work was to find a way of regaining the periodicity in Mendeleev's table through the comparison of density functions.

The purpose of this contribution is to give an overview of the results which center around the atomic density function and the recovery of the periodicity. Since all the calculations are based on atomic density functions, it is appropriate to revisit the construction of these densities in some depth. First a workable definition of the density function is established in the framework of the multi-configuration Hartree--Fock method (MCHF)~\cite{Fischer:2007hi} and the spherical harmonic content of the density function is discussed. A spherical density function is established in a natural way, by using spherical tensor operators. The proposed expression can be evaluated for any multi-configuration state function corresponding to an atom in a particular well-defined state and a recently developed extension of the MCHF code~\cite{Borgoo2010426} is used for that purpose. Three illustrative examples are given. In the next section relativistic density functions for the relativistic Dirac--Hartree--Fock method~\cite{DESCLAUX:1975} are defined. The latter will be used for a thorough analysis of the influence of relativistic effects on electron density functions later on in this paper.

The analysis of atomic density functions can be furthered by comparing them in pairs. Specifically, the use of quantum similarity measures and indices as defined by Carb\'o~\cite{CARBO:1980} has shown that particular influences on the density functions can be estimated in this way. Here this feature is demonstrated by reviewing three case studies: i) the $LS$-term dependence of Hartree--Fock densities, ii) the comparison of atoms throughout the periodic table~\cite{Borgoo:2004} and iii) the quantitative evaluation of the influence of relativistic effects, via a comparison of non-relativistic Hartree--Fock densities with Dirac--Hartree--Fock relativistic densities~\cite{Borgoo:2007ad}.

In the final part of this contribution, information theory is introduced. After a brief revision of the relevant concepts, a functional based on the Kullback--Leibler measure~\cite{Kullback:Leibler:1951} is constructed for the investigation of atomic density functions throughout Mendeleev's table. Since the quantum similarity does not reveal the expected periodic patterns, it is significant to show that it is actually possible to regain the periodicity by constructing an appropriate functional~\cite{Borgoo:2004}. By considering the integrand of the Kullback--Leibler measure and comparing it locally for two atoms, a quantum similarity measure can be constructed which does reveal periodic patterns~\cite{Borgoo:2007ad}.

\section{The multi-configuration many-electron wave function} \label{secsec:2}

The total energy which results from the Hartree--Fock equations is due to electrons moving independently in an averaged, central potential. Any improvements to the energy (still in the context of the non-relativistic Schr\"odinger equation) are said to be due to correlation effects as a direct consequence of the electron--electron interaction. In fact it is common to define the correlation energy as
\begin{equation}
E_{\mbox{\sc corr}} = E_{\mbox{\sc exact}} - E_{\mbox{\sc hf}} \, ,
\end{equation}
where $E_{\mbox{\sc exact}}$ is the exact non-relativistic energy and $E_{\mbox{\sc hf}}$ the energy due to the solutions of the corresponding Hartree--Fock equations.

A particularly natural way to improve the Hartree--Fock energy -- i.e. include correlation energy -- is by departing from the single-configuration approximation. In Hartree--Fock calculations a rigid orbital picture is assumed, where electrons have a fixed place in a given electron configuration. By allowing electrons to occupy different orbitals and allowing several electron configurations, the variational approach can be applied on a significantly larger set of trial wave functions.

In the multi-configuration Hartree--Fock (MCHF) approach, the $N$-electron wave function $\Psi_{\alpha L S M_L M_S}$
is a linear combination of $M$ configuration state functions (CSFs) $\Phi_{\alpha_i LS M_L M_S}$ which are eigenfunctions of the total angular momentum $L^2$, 
the spin momentum $S^2$ and their projections $L_z$ and $S_z$,   with eigenvalues 
$\hbar^2 L(L+1) $, $  \hbar^2 S(S+1) $, $ \hbar M_L $ and $ \hbar M_S $, respectively

\begin{equation}
\label{MCHF_exp}
\Psi_{\alpha LSM_LM_S}(\vec{x}_1, \cdots \vec{x}_N) 
= 
\sum_{i=1}^M \; c_i \; \Phi (\alpha_i L S M_L M_S; \vec{x}_1, \cdots \vec{x}_N) \; ,
\end{equation}
with 
\begin{equation}
\sum_{i=1}^M |c_i|^2 = 1 \; .
\end{equation}

The mixing coefficients $\{ c_i \}$ and the radial functions $\{ R_{n_il_i} (r) \}$, constituting the one-electron basis, are solutions of the multi-configuration Hartree--Fock 
method in the MCHF approach. For a given set of orbitals, the mixing coefficients may also be the solution of 
the configuration interaction (CI) problem. The relativistic corrections can be taken into account by diagonalizing the Breit-Pauli Hamiltonian 
\cite{HIBBERT:1991yb} in the $LSJ$-coupled CSF basis to get the intermediate coupling eigenvectors

\begin{equation}
\label{MCHF_BP_exp}
\Psi_{\alpha J M}(\vec{x}_1, \cdots \vec{x}_N) 
= 
\sum_{i=1}^{M} \; a_i \; \Phi (\alpha_i L_i S_i  J M ; \vec{x}_1, \cdots \vec{x}_N) \; ,
\end{equation}
with 
\begin{equation}
\sum_{i=1}^M |a_i|^2 = 1 \; .
\end{equation}

In the MCHF approach, the trial wave functions are of the form~\rref{MCHF_exp} and the energy expression for a given state becomes 
\begin{equation} \label{Theory_MCHFenergy}
E(\alpha LS) = \sum_i^M \sum_j^M c^*_i c_j H_{ij} \; ,
\end{equation}
where 
\begin{equation}
H_{ij} \equiv \langle \Phi_{\alpha_i L S M_L M_S} | H | \Phi_{\alpha_j L S M_L M_S} \rangle = \sum_{ab} q_{ab}^{ij} I_{ab} + \sum _{abcd; k} v_{abcd;k}^{ij}R^k (ab,cd) \; ,
\end{equation}
where $a \equiv (n_al_a)$ and $b \equiv (n_b l_b)$ and the sums run over all occupied orbitals in the respective configuration $i$ and $j$.

In the MCHF context the variational principle is applied to the energy functional in equation~\rref{Theory_MCHFenergy}. A stationary solution is obtained by minimizing the energy with respect to variations in the radial wave functions $P_{nl} (r) \equiv r R_{nl}(r)$ satisfying the orthonormality conditions 
\begin{equation}
N_{nl,n'l} \equiv \int_0^{\infty} P_{nl} (r) P_{n'l} (r) \; dr = \delta_{nn'}
\end{equation}
and 
\begin{equation}
\sum_{i=1}^M c_i^2 = 1 \; .
\end{equation}

The energy expression~\rref{Theory_MCHFenergy} can be written in matrix notation
\begin{equation}
E(\alpha LS) = \mathbf{C}^t \mathbf{H} \mathbf{C} \; ,
\end{equation}
where $\mathbf{H}  \equiv \left( H_{ij} \right)$, $\mathbf{C}$ is the column matrix of the expansion coefficients and $\mathbf{C}^t$ its transpose.

This gives the eigenvalue problem for the expansion coefficients
\begin{equation}
\mathbf{H}\mathbf{C} = \mathbf{C}E \; .
\end{equation}

To obtain a self consistent field solution to the MCHF problem, two optimizations need to be performed i.e. one for the variation of the one-electron radial orbitals $\{P_{nl} (r)\}$ in the wave function and one for the expansion coefficients. This can be done by consecutively iterating, first the orbital optimization followed by the coefficient optimization.

\section{\label{sec2}On the symmetry of the density function}

In the first part of this paragraph we review the study of the spherical harmonic content of the density function for atoms in a well defined state~\rref{MCHF_exp} or~\rref{MCHF_BP_exp}. The spherical density functions, which reveal the familiar shell structure, are discussed and illustrative examples are given.

\subsection{The non-spherical density function }

The so-called ``generalized density function"  \cite{McWeeny:1992dz} or the ``first order reduced density matrix"  
\cite{Helgaker:2000ir} is  a special case of the reduced density matrix  \cite{DAVIDSON:1976cs,McWeeny:1992dz} 
\begin{equation}
\label{gamma_1}
\gamma_1(\vec{x}_1,\vec{x'}_1) = 
N \int \Psi(\vec{x}_1,\vec{x}_2, \ldots , \vec{x}_N)  \; \Psi^*(\vec{x'}_1,\vec{x}_2, \ldots , \vec{x}_N) \; d\vec{x}_2 \ldots d \vec{x}_N \; ,
\end{equation}
where $\Psi(\vec{x}_1,\vec{x}_2, \ldots , \vec{x}_N)$ is the total wave function of an $N$ electron system and 
$\Psi^* (\vec{x}_1,\vec{x}_2, \ldots , \vec{x}_N)$ is 
its complex conjugate. The spin-less total electron density function $\rho(\vec{r})$ is defined as the first 
order reduced density matrix, integrated over the spin and evaluated for $\vec{x}_1 = \vec{x'}_1$
\begin{equation}
\rho(\vec{r}_1) = \int \gamma_1(\vec{x}_1,\vec{x}_1) \; d\sigma_1\; . \label{density}
\end{equation}
This electron density function is normalized to the number of electrons of the system
\begin{equation}
\int \rho(\vec{r})  \; d \vec{r} =  \int \rho(\vec{r}) \;  r^2  \sin\vartheta \; dr d\vartheta d\varphi = N \; .
\end{equation}
As discussed in \cite{Helgaker:2000ir}, the single particle density function can be calculated by evaluating the expectation value of the 
$ \delta (\vec{r}) $ operator, 
\begin{equation}
\label{xpectation_value}
\rho(\vec{r}) = 
\int \Psi(\vec{x}_1,\vec{x}_2, \ldots , \vec{x}_N) \;
\delta (\vec{r}) \;
\Psi^*(\vec{x}_1,\vec{x}_2, \ldots , \vec{x}_N) \; d\vec{x}_1 d\vec{x}_2 \ldots d \vec{x}_N \; ,
\end{equation}
where  $ \delta (\vec{r}) $ probes the presence of electrons at a particular point in space and can be written as the one-electron first-quantization operator
\begin{equation} 
\label{firstquant}
\delta (\vec{r}) = \sum_{i=1}^{N}\delta (\vec{r} - \vec{r}_i) \, .
\end{equation}
The exact spin-less total electron density function (\ref{xpectation_value}) evaluated for an eigenstate with well-defined quantum numbers $(L S M_L M_S)$

\begin{equation} 
\label{exact_density}
\rho(\vec{r}) ^{LSM_LM_S} 
=   \sum_{lm} Y_{lm} (\vartheta, \varphi) \frac{1}{r^2 } \; 
\langle \Psi_{\alpha LSM_LM_S} \vert     
\sum_{i=1}^{N}  \delta (r - r_i) \;   Y_{lm} ^\ast (\vartheta_i, \varphi_i)  
\vert \Psi_{\alpha LSM_LM_S} \rangle \; ,
\end{equation} 
becomes
\begin{eqnarray} 
\label{exact_density_2}
\rho(\vec{r}) ^{LSM_LM_S}
& = & \sum_{l=0}^{L} \rho (r) ^{LSM_L M_S}_{2l} Y_{2l \; 0} (\vartheta, \varphi) \; ,
\end{eqnarray}
where 
\begin{eqnarray}
\label{rho_2l}
\rho (r) ^{LSM_L M_S}_{2l} &=& 
\frac{1}{r^2 } \; 
(-1)^{L - M_L} 
\left( \begin{array}{ccc} L & 2l & L \\ -M_L  & 0 & M_L \end{array} \right) \nonumber \\
&\times& \langle \Psi_{\alpha LS } \Vert     \sum_{i=1}^{N}  \delta (r - r_i) \;   Y_{2l} ^\ast (\vartheta_i, \varphi_i) \Vert \Psi_{\alpha LS} \rangle \; .
\end{eqnarray}
This result, which can be found by applying the Wigner--Eckart theorem~\cite{Cowan:1981xr}, recovers Fertig and Kohn's analysis~\cite{Fertig:2000pt} for the density corresponding to a 
well-defined $(LSM_LM_S)$ eigenstate of the Schr\"{o}dinger equation. 
The spherical harmonic components in the density are limited to $l$-even contributions, because the bra and the ket need to be of the same parity $\pi = (-1)^{\sum_i l_i}$.

In this paper, the authors
observed that the self-consistent field densities obtained 
via the Hartree and Hartree--Fock methods generally violate the specific finite spherical harmonic 
content of $\rho(\vec{r}) ^{LSM_LM_S}$.
They also mention that this exact form can be obtained by 
spherically averaging the effective potential, yielding single-particle states with good angular 
momentum quantum numbers. 
The atomic  structure software package {\sc atsp2K}~\cite{Fischer:2007hi} applies this approach, as was done in the original 
atomic Hartree--Fock theory \cite{Slater:1930yu,Hartree:1957zf,Fischer:1977dz}. This implies two things: 
i) the density function $\rho (\vec{r})^{LSM_LM_S}$ calculated from any multiconfiguration wave function 
of the form (\ref{MCHF_exp}), is not {\em a priori} spherically symmetric, 
ii) this density function will contain all spherical harmonic components (up to $2L$) as long as 
the one-electron orbital active set spanning the configuration space is $l$-rich enough. 

The density function can also be expressed in second quantization \cite{McWeeny:1992dz}. Introducing the notation 
$ q \equiv n_q l_q m_{l_q} m_{s_q} $ for spin-orbitals, expression~(\ref{gamma_1}) becomes 
\begin{equation}
\gamma_1(\vec{x}_1,\vec{x'}_1) = \sum_{pq} D_{pq}  \; \psi^*_p(\vec{x'}_1) \psi_q(\vec{x}_1) \label{gamma} \; ,
\end{equation}
where $D_{pq}$ are elements of the density matrix which are given by
\begin{equation}
D_{pq} \equiv 
\braket< \Psi |a^\dag_p a_q|\Psi> \; .
\end{equation}
The sum in eq.~(\ref{gamma}) runs over all possible pairs of quartets of quantum numbers 
$p$ and $q$. The spin-less density function~(\ref{density}) calculated from 
$ \rho(\vec{r}) =  \braket  < \Psi \vert \hat{\delta}(\vec{r})  \vert \Psi >$, 
using the second quantized form of the operator 
\begin{eqnarray}
\label{secont_quant_1}
\hat{\delta}(\vec{r}) & \equiv & \sum_{pq} a^\dag_p a_q \; \delta_{m_{s_p},m_{s_q}} 
\langle \psi_p(\vec{r'}) 
\vert   \frac{1}{r^2 \sin \vartheta} \; \delta (r - r') \; \delta (\vartheta - \vartheta') \; \delta (\varphi - \varphi')  \vert 
\psi_q(\vec{r'}) \rangle   
\nonumber \\
& = & \sum_{pq} a^\dag_p a_q \; \delta_{m_{s_p},m_{s_q}} R^\ast_{n_p l_p}(r) Y^\ast_{l_p m_{l_p}}(\vartheta,\varphi) 
R_{n_q l_q}(r)    Y_{l_q m_{l_q}}(\vartheta,\varphi) \; ,
\end{eqnarray}
yields
\begin{equation}
\label{rho_r}
\rho(\vec{r}) =  
\sum_{pq}    D_{pq} \; \delta_{m_{s_p},m_{s_q}} \; 
R^\ast_{n_p l_p}(r)Y^*_{l_p m_{l_p}}(\vartheta,\varphi) R_{n_q l_q}(r)Y_{l_q m_{l_q}}(\vartheta,\varphi) \; .
\end{equation}

To illustrate the spherical harmonics content of the density in the Hartree--Fock approximation, 
consider the atomic term $ 1s^2 2p^2 (\; ^3P) 3d \; ^4F$ for which the $(M_L,M_S) = (+3,+3/2)$ subspace reduces to a single Slater determinant 
\begin{equation}
\Psi_{\alpha L S M_L M_S} = 
\Phi (1s^2 2p^2 (\; ^3P) 3d \; ^4F_{+3,+3/2} )   =
|1s \overline{1s} 2p_{+1} 2p_{0} 3d_{+2} | \; .
\end{equation}
When evaluating (\ref{rho_r}), all non-zero $D_{pq}$-values appear on the diagonal ($p=q$), yielding 
\begin{equation}
\label{not_spher}
\rho(\vec{r})^{^4F_{+3,+3/2}} =
|\psi_{1s}(\vec{r}) |^2 + |\psi_{\overline{1s}} (\vec{r}) |^2 + |\psi_{2p_{+1}} (\vec{r})|^2 + |\psi_{2p_0} (\vec{r})|^2 + |\psi_{3d_{+2}} (\vec{r})|^2
\; .
\end{equation}
This density has a clear {\em non}-spherical angular dependence.
However, referring to~\cite{Varshalovich:yg}
\begin{equation}
W^{\parallel}_{JM} (\vartheta) \equiv \vert Y_{JM}(\vartheta,\varphi) \vert^2 
= \sum_{n=0}^J b_n (J,M) \; P_{2n} (\cos \vartheta )
= \sum_{n=0}^J b'_n (J,M) \; Y_{2n \; 0} (\vartheta,\varphi)
\end{equation}
one recovers the even Legendre polynomial content of the density, although not 
reaching the  $(2L = 6)$ limit $Y_{6 \; 0}(\vartheta, \varphi) $ of 
the exact density (\ref{exact_density_2}). This limit will be attained  
when extending the one-electron orbital active set to higher angular momentum values for building 
a correlated wave function. 

As discussed in detail in~\cite{Borgoo2010426}, contributions to the density function corresponding to $(p \neq q)$ can appear through off-diagonal matrix elements in the CSF basis. These contributions will be present for parity conserving single electron excitations of the type $\vert l_1 q \rangle \rightarrow \vert l_2 q \rangle$.

The ``offending" spherical harmonic contributions described by Fertig and Kohn \cite{Fertig:2000pt} do not occur in the MCHF calculation of the density function, whatever the maximum $l$-value of the orbital active space~\cite{Borgoo2010426}.

\subsection{\label{secsec:3}The spherical density function }

A {\em spherically} symmetric density function can be obtained for an arbitrary  CSF 
$\Phi_{\alpha LS M_L M_S}$ by averaging the  $(2L+1)(2S+1)$
magnetic components of the spin-less density function 
\begin{equation} \label{av}
\rho(\vec{r}) ^{LS} \equiv \frac{1}{(2L+1)(2S+1)} \; \sum_{M_L M_S} \rho(\vec{r}) ^{LSM_LM_S} \; ,
\end{equation}
where $ \rho(\vec{r}) ^{LSM_LM_S} $ is constructed according to eq.~(\ref{rho_r})
\begin{equation}
\label{rho_r_LS}
\rho(\vec{r}) ^{LSM_LM_S} = \sum_{pq} \braket< \Phi_{\alpha LS M_L M_S} |a^\dag_{p} a_{q}|
\Phi_{\alpha LS M_L M_S} >  \;  \delta_{m_{s_p},m_{s_q}} \; \psi^*_p(\vec{r}) \psi_q(\vec{r}) \; .
\end{equation}

Applying equations~(\ref{av}) and (\ref{rho_r_LS})
for the atomic term $ 1s^2 2p^2 (\; ^3P) 3d \; ^4F$ considered in the previous section, 
we simply get
\begin{equation}
\rho(\vec{r}) ^{^4F} = \frac{1}{4 \pi r^2}  \left\{
2 P^2_{1s}(r)  + 2 P^2_{2p}(r) + P^2_{3d}(r) \right\} \; ,
\end{equation}

\noindent which is, {\em in contrast} to eq.~(\ref{not_spher}), obviously spherically symmetric.
The sum over $(M_L,M_S)$ performed in (\ref{av}) 
guarantees, for any $nl$-subshell, the presence of all necessary components $ \{ Y_{lm_l} \; \vert \; m_l = -l, \ldots +l \} $ 
with the  same weight factor,   which permits the application of Uns\"{o}ld's theorem \cite{Unsold:1927tn}
\begin{equation}
\label{Unsold}
\sum_{m_l=-l}^{+l} \; \vert Y_{lm_l} (\vartheta, \varphi) \vert ^2 = \frac{2l+1}{4 \pi} 
\end{equation}
and yields the spherical symmetry.  This result  is valid for any single CSF 
\begin{equation}
\label{gen}
\rho(\vec{r}) ^{LS} = \frac{1}{4 \pi r^2}  \sum_{nl} q_{nl}  P^2_{nl}(r)  \; ,
\end{equation}
where $q_{nl}$ is the occupation number of $n l$-subshell. Its sphericity explicitly appears by rewriting (\ref{gen}) as
\begin{equation}
\label{rho_r_vec}
\rho(\vec{r})  = \rho(r) \; \vert Y_{00} (\vartheta, \varphi) \vert ^2 
= \frac{D(r)}{r^2} \; \vert Y_{00} (\vartheta, \varphi) \vert ^2   \; ,
\end{equation}
with
\begin{equation}
\label{rho_r_scal}
\rho(r) \equiv \frac{1}{r^2} \sum_{nl} q_{nl}   P^2_{nl}(r) \; ,
\end{equation}
and 
\begin{equation}
\label{D_r}
D(r) \equiv r^2 \rho(r) = \sum_{nl} q_{nl}   P^2_{nl}(r)  = 
\sum_{nl} q_{nl} \; r^2  R^2_{nl}(r) \; .
\end{equation}\\
The {\em radial distribution} function $D(r)$ represents the probability of finding an electron 
between the distances $r$ and $r + dr$ from the nucleus, regardless of direction
.
This radial density function reveals the 
atomic shell structure when plotted as function of $r$. Its integration over $r$ gives the total number of electrons of the system
\begin{equation}
\label{integral_density}
\int_{0}^\infty D(r)  \; d r = \int_{0}^\infty r^2 \rho(r)  \; d r =  \sum_{nl} q_{nl} = N \; .
\end{equation}

Where above the spherical symmetry of the average density (\ref{av}) is  demonstrated for a single CSF thanks to Uns\"{o}ld's theorem, it can be
demonstrated in the general case by combining (\ref{av}), (\ref{exact_density_2}) 
and the 3-$j$ sum rule \cite{Cowan:1981xr}
\begin{equation}
\sum_{M_L} (-1)^{L - M_L} 
\left( \begin{array}{ccc} L & k & L \\ -M_L  & 0 & M_L \end{array} 
 \right) =  (2k + 1)^{1/2} \; \delta_{k,0}
\end{equation}
for each $k = 2l$ contribution~(\ref{rho_2l}). However, the radial density $\rho(r)$ will be more complicated than (\ref{rho_r_scal}), 
involving mixed contributions of the type 
$P_{n'l}(r) P_{nl}(r) = r^2 R_{n'l}(r) R_{nl}(r)$, as developed below.

Instead of obtaining a spherically symmetric density function by averaging the magnetic components $ \rho(\vec{r}) ^{LSM_LM_S}$ through
eq.~({\ref{av}), one can build a radial density operator associated to the function (\ref{D_r})  
which is spin- and angular-independent, i.e. independent of the spin~($\sigma$) and 
angular ($\vartheta, \varphi$) variables. Adopting the methodology used by 
Helgaker {\em et al} \cite{Helgaker:2000ir} for defining the spin-less density operator, 
we write a general first quantization spin-free {\it radial} operator 
\begin{equation}
\label{oneparticle_FQ}
f = \sum_{i=1}^N f(r_i)
\end{equation}
in second quantization as
\begin{equation}
\label{oneparticle_SQ}
\hat{f} = \sum_{pq} f_{pq}  \; a_p^\dag a_q   \; ,
\end{equation}
where $f_{pq}$ is the one-electron integral
\begin{equation}
f_{pq} = \int \psi^*_p(\vec{x}) f(r) \psi_q(\vec{x}) r^2 \sin \vartheta  dr d\vartheta d\varphi d\sigma
\; .
\end{equation}
Applying this formalism to the radial density operator
\begin{equation} 
\label{first_quant_delta_r}
\delta (r) \equiv \sum_{i=1}^{N}\delta (r - r_i) \, ,
\end{equation}
and  using the spin-orbital factorization for both $p$ and $q$ quartets, we obtain the
second quantization form 
\begin{equation}
\label{second_quant_delta_r}
\hat{\delta}(r) = \sum_{pq} d_{pq}(r)  \, a_p^\dag a_q \; ,
\end{equation}
with
\begin{equation} 
\label{delta_r_pq}
d_{pq}(r) =
\delta_{l_p l_q} \; \delta_{m_{l_p} m_{l_q} } \; \delta_{m_{s_p} m_{s_q}}  \; R^*_{n_pl_p} (r) R_{n_ql_q} (r) r^2
\; ,
\end{equation}
where the Kronecker delta arises from the orthonormality property of the spherical harmonics and spin functions.
With real radial one-electron functions, the operator (\ref{second_quant_delta_r}) becomes
\begin{equation}
\label{second_quant_delta_r_2}
\hat{\delta}(r) = 
\sum_{n', l', m_l', m_s', n, l, m_l, m_s, } 
\delta_{l' l} \; \delta_{m'_l m_l} \; \delta_{m'_s m_s} \; a^\dag_{n'l'm'_lm'_s} a_{nlm_lm_s} \;
R_{n'l'} (r) R_{nl} (r) r^2 
\end{equation}
\begin{equation}
= \sum_{n',n}  \sum_{l, m_l, m_s}  \; a^\dag_{n'lm_lm_s} a_{nlm_lm_s} \; R_{n'l} (r) R_{nl} (r) r^2  \; .
\end{equation}
Its expectation value  provides the radial density function 
$D(r) = r^2 \rho(r) = 4 \pi r^2 \rho( \vec{r})$ defined by (\ref{rho_r_vec}) 
and (\ref{D_r}).

Building the coupled tensor of ranks $(00)$   from the $[2(2l+1)]$ components of the 
creation and annihilation operators \cite{Judd:1967xh}
\begin{equation}
\label{coupled_tensor_00}
\left(
{\bf a}^\dag_{n'l} {\bf a}_{nl} \right) ^{(00)}_{00}
= -\frac{1}{\sqrt{2(2l+1)}} \; \sum_{m_l m_s} 
a^\dag_{n'lm_lm_s} a_{nlm_lm_s}   \; ,
\end{equation}
the operator (\ref{second_quant_delta_r_2}) becomes
\begin{equation}
\label{second_quant_delta_r_3}
\hat{\delta}(r) = - \sum_l \sqrt{2(2l+1)} 
\sum_{n',n}  \; \left( {\bf a}^\dag_{n'l} {\bf a}_{nl} \right) ^{(00)}_{00} 
\; R_{n'l} (r) R_{nl} (r) r^2  \; .
\end{equation}
The expectation value of this operator provides the spherical density function for any atomic state.
Note that, in contrast to (\ref{rho_r_LS}), the tensorial ranks (00) garantee the diagonal character in $L, S, M_L $ and $M_S$, 
thanks to Wigner-Eckart theorem 
\begin{eqnarray}
\label{WE_00}
\langle \alpha L S M_L  M_S \vert T^{(00)}_{00} \vert \alpha' L' S' M_L'  M_S' \rangle
&=& (-1)^{L + S - M_L - M_S}  \nonumber \\
&\times& \left( \begin{array}{ccc} L & 0 & L' \\ -M_L  & 0 & M_L' \end{array}    \right)
  \left( \begin{array}{ccc} S & 0 & S' \\ -M_S  & 0 & M_S' \end{array}    \right) \nonumber \\
&\times&  \langle \alpha L S \Vert T^{(00)} \Vert \alpha' L'  S'  \rangle\; .
\end{eqnarray}
Moreover, the $M_L / M_S$ independence emerges from the special $3j$-symbol
\begin{equation}
\label{special_3j}
\left( \begin{array}{ccc} j & 0 & j' \\ -m_j  & 0 & m_j' \end{array}    \right)
= (-1)^{j-m} (2j + 1)^{-1/2} \delta_{jj'} \delta_{m{_j} m_{j}'} \; .
\end{equation}
In other words, where the non-spherical components are washed out by the averaging process (\ref{av}), 
they simply do not exist and will never appear for the 
density calculated from (\ref{second_quant_delta_r_3}), for any $(M_L, M_S)$ magnetic component.


The radial distribution function $D(r) \equiv r^2 \rho(r)$ can be calculated from the expectation value of the operator (\ref{second_quant_delta_r_3}),
using the wave function (\ref{MCHF_exp}) or (\ref{MCHF_BP_exp}). In the most general case (expansion (\ref{MCHF_BP_exp})), using the $(LS)J$-coupled form of the
excitation operator, 
\begin{equation}
\left( {\bf a}^\dag_{n'l} {\bf a}_{nl} \right) ^{(00)0}_{0}  = 
\left( {\bf a}^\dag_{n'l} {\bf a}_{nl} \right) ^{(00)}_{00}  \; ,
\end{equation}
one obtains
\begin{equation}
\label{eq:density_BP}
\langle \Psi_{\alpha J M } \vert  \hat{\delta}(r) \vert \Psi_{\alpha J M } \rangle
= (-1)^{J - M} 
\left( \begin{array}{ccc} J & 0 & J \\ -M  & 0 & M \end{array}    \right)
\langle \Psi_{\alpha  J }  \| \widehat{F}^{(00)0}_{\rho} \| \Psi_{\alpha J } \rangle
\end{equation}
with
\begin{equation}
\label{eq:F_00_LSJ}
\widehat{F}^{(00)0}_{\rho, 0} \; = \;
    - {\sum_{l = 1}} \sqrt{2 \left( 2l+1 \right)}
 \; {\sum_{n,n^{\prime}}}
    \left( {\bf a}^\dag_{n'l} {\bf a}_{nl} \right) ^{(00)0} _0
\;  I_{\rho} \left( n^{\prime}l, nl \right)  \; ,
\end{equation}
and 
\begin{equation}
\label{eq:I_rho}
I_{\rho}\left(n^{\prime}l, nl \right) (r) \; \equiv \;
R_{n'l} (r) R_{nl} (r) r^2 \; .
\end{equation}
The diagonal reduced matrix element (RME) evaluated with the Breit-Pauli eigenvector (\ref{MCHF_BP_exp}) has the following form
\begin{equation}
\label{eq:density_rme_BP}
\langle \Psi_{\alpha J } \| \widehat{F}^{(00)0}_{\rho} \| \Psi_{\alpha J } \rangle
= \sum_{i,j} a^\ast_i a_j \; 
\langle \Phi (\alpha_i L_i S_i  J )  \| \widehat{F}^{(00)0}_{\rho} \| \Phi (\alpha_j L_j S_j  J ) \rangle
\end{equation}
where the RME in the $(LS)J$ coupled basis reduces to
\begin{eqnarray}
\label{eq:density_rme_CSF}
\lefteqn{
\langle \Phi (\alpha_i L_i S_i  J M)  \| \widehat{F}^{(00)0}_{\rho} \| \Phi (\alpha_j L_j S_j  J M) \rangle}  \nonumber \\
&=&
\sqrt{\frac{2J+1}{(2L_i+1)(2S_i+1)}} 
 \delta_{L_i,L_j} 
 \delta_{S_i,S_j} \nonumber \\
&\times& \langle \Phi (\alpha_i L_i S_i )  \| \widehat{F}^{(00)}_{\rho} \| \Phi (\alpha_j L_j S_j  ) \rangle
\end{eqnarray}
and
\begin{equation}
\label{eq:Density_operator_SC}
\widehat{F}^{(00)}_{\rho, 0 0} \; = \;
    - {\sum_{l = 1}} \sqrt{2 \left( 2l+1 \right)}
 \; {\sum_{n,n^{\prime}}}
    \left( {\bf a}^\dag_{n'l} {\bf a}_{nl} \right) ^{(00)}_{00} 
\;  I_{\rho} \left( n^{\prime}l, nl \right)  \; .
\end{equation}

From a comparison of the operator (\ref{eq:Density_operator_SC}) with the non-relativistic one-body Hamiltonian operator (see eq.~(A5) of \cite{OLSEN:1995pl}),
one observes that the angular coefficients of the radial functions $I_{\rho}\left(n^{\prime}l, nl \right) (r)$ 
are identical to those of the one-electron Hamiltonian radial integrals
$I_{n^{\prime}l, nl}$ ,  as anticipated from McWeeny analysis \cite{McWeeny:1992dz}. 
These angular coefficients can be derived by working out the  matrix elements of a one--particle 
scalar operator $\widehat{F}_{\rho}^{(00)}$ between configuration state functions with $u$
open shells, as explicitly derived by Gaigalas {\em et al} \cite{Gaigalas:2001oe}.

\section{Three tangible examples} \label{Density_Applications}

First the evaluation of the density function and the influence of correlation effects is illustrated by plotting in figure \ref{fig_Be} the radial density distribution $D(r) =  r^2 \rho(r) $ for a CAS-MCHF wave function of the beryllium ground state (Be $1s^2 2s^2 \; ^1S$), using a $n=9$ orbital active set. 
In the same figure, the Hartree--Fock radial density is compared with the one obtained with two correlation models: i) the $n=2$ CAS-MCHF expansion, largely dominated by the near-degeneracy mixing associated to the Layzer complex $1s^2 \{2s^2 + 2p^2 \}$ and ii) the $n=9$ CAS-MCHF. From the plotted results we notice that  the density of the $n=2$ calculation already contains the major correlation effects, compared to the $n=9$ calculation. Indeed, the density does not seem to change a lot by going from the $n=2$ to the $n=9$ orbital basis, the valence double excitation $1s^2 2p^2$ contributing for 9.7\% of the wave function.  From the energy point of view however, this observation is somewhat surprising (see Table~\ref{table_BE}): the correlation energy associated to the $n=2$ CAS-MCHF solution ``only'' represents  47\% of the $n=9$ correlation energy.

\begin{figure}
\begin{center}
\resizebox{70mm}{!}{\includegraphics[viewport=40 400 500 800,angle=0]{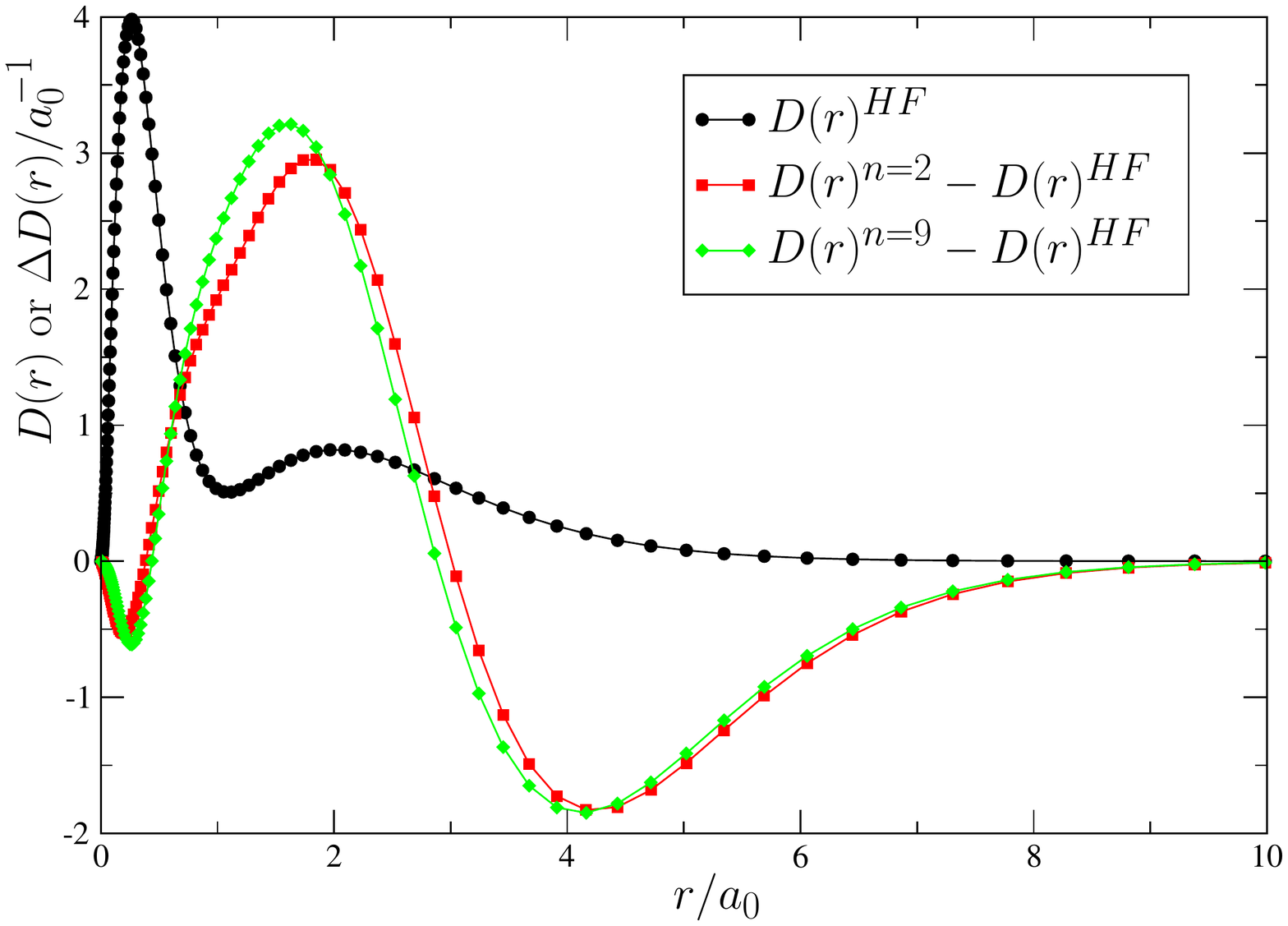}}
\caption{Density of $^1$S Be ground state for different CAS-MCHF wave function as compared to Hartree--Fock (HF). Density differences have been scaled by a factor 100.}
\label{fig_Be}
\end{center}
\end{figure}

\begin{table}
\begin{center}
\begin{tabular}{|c|c|c|}
\hline
model & energy (a.u.) & correlation energy (a.u.) \\
\hline
HF  &  -14.573~023 & \\ 
$n=2$-CAS & -14.616~856 & $E_{corr}^{n=2} = E^{n=2} - E^{HF} = 0.043~832$\\ 
$n=9$-CAS & -14.667~013 & $E_{corr}^{n=9} = E^{n=9} - E^{HF} = 0.093~986$\\ 
 \hline
\end{tabular}
\end{center}
\caption{Total energy for the ground state of Be with different correlation models.}
\label{table_BE}
\end{table}

As a second example,  we illustrate the influence of relativistic effects -- in the Breit-Pauli approximation -- on the density function of the Be-like O$^{4+}$ atom, by comparing the densities of the fine-structure states $1s^2 2s 2p \; ^3P^\circ_0$, $^3P^\circ_1$ and $^3P^\circ_2$. From the plots in Figure~\ref{fig_BP} and the data given in Table~\ref{table_BP} we observe that the largest energy difference corresponds to the largest difference in density function. More bound is the level, higher is the electron density in the inner region.
The influence of relativistic effects on the density function will be discussed thoroughly below.

\begin{table}
\begin{center}
\begin{tabular}{|r|c|c|}
\hline
model & energy (a.u.) & energy difference (a.u.)\\
\hline
$1s^2 2s2p \; ^3P^o_0$ & -68.032~086 & \\ 
$^3P^o_1$ & -68.031~473 &$E(\; ^3P_1 - \; ^3P_0) = 0.000~613$\\ 
$^3P^o_2$ & -68.030~102 &$E(\; ^3P_2 - \; ^3P_1) = 0.001~370$\\ 
 \hline
\end{tabular}
\end{center}
\caption{Total energy for 2s2p $^3$P$^{\mathbf{\tiny o}}_{\mbox{\tiny J}}$ fine-structure levels of O$^{4+}$}
\label{table_BP}
\end{table}

Finally a third example is given that is relevant when studying the electron affinities, as it is often interesting to investigate the differential correlation effects between the negative ion and the neutral system~\cite{PhysRevA.81.042522}. Figure~\ref{fig_S} displays the density functions $D(r)$ of both the [Ne]$3s^2 3p^4 \; ^3P$ ground state of neutral Sulphur (S) and the [Ne]$3s^2 3p^5 \; ^2P^\circ$ ground state of the negative ion S$^-$, evaluated with elaborate correlation models~\cite{PhysRevA.81.042522}, together with their difference $\Delta D(r)$. The latter integrates to unity and reveals that the ``extra'' electron resides in the valence area.

\begin{figure}
\begin{center}
\resizebox{105mm}{!}{\includegraphics[clip,angle=-90]{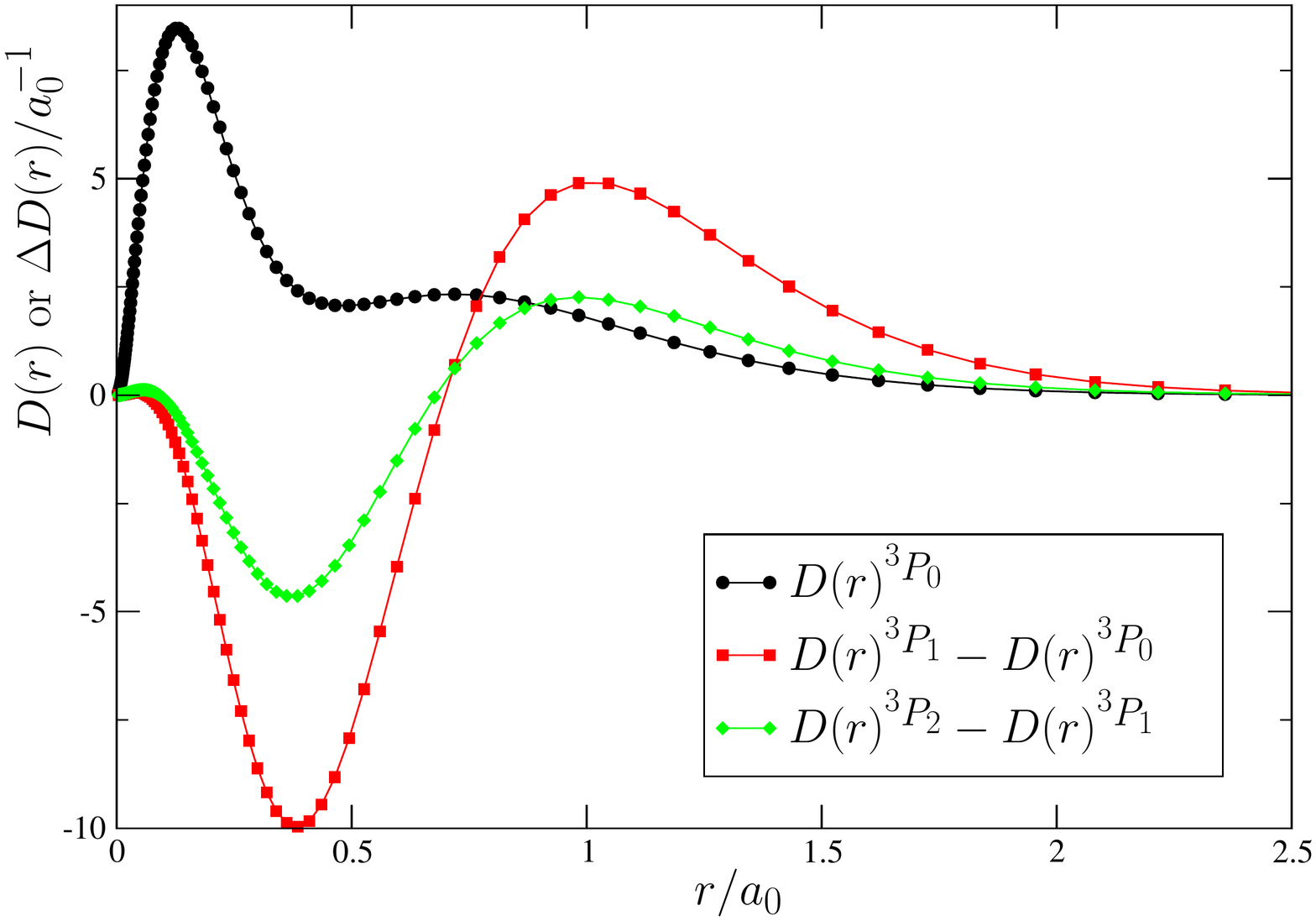}}
\caption{Comparison of the $^3$P$_0$, $^3$P$_1$ and $^3$P$_2$ radial density functions for O$^{4+}$. Density differences have been scaled by a factor 10~000.}
\label{fig_BP}
\vspace{-5mm}
\resizebox{105mm}{!}{\includegraphics[clip,angle=-90]{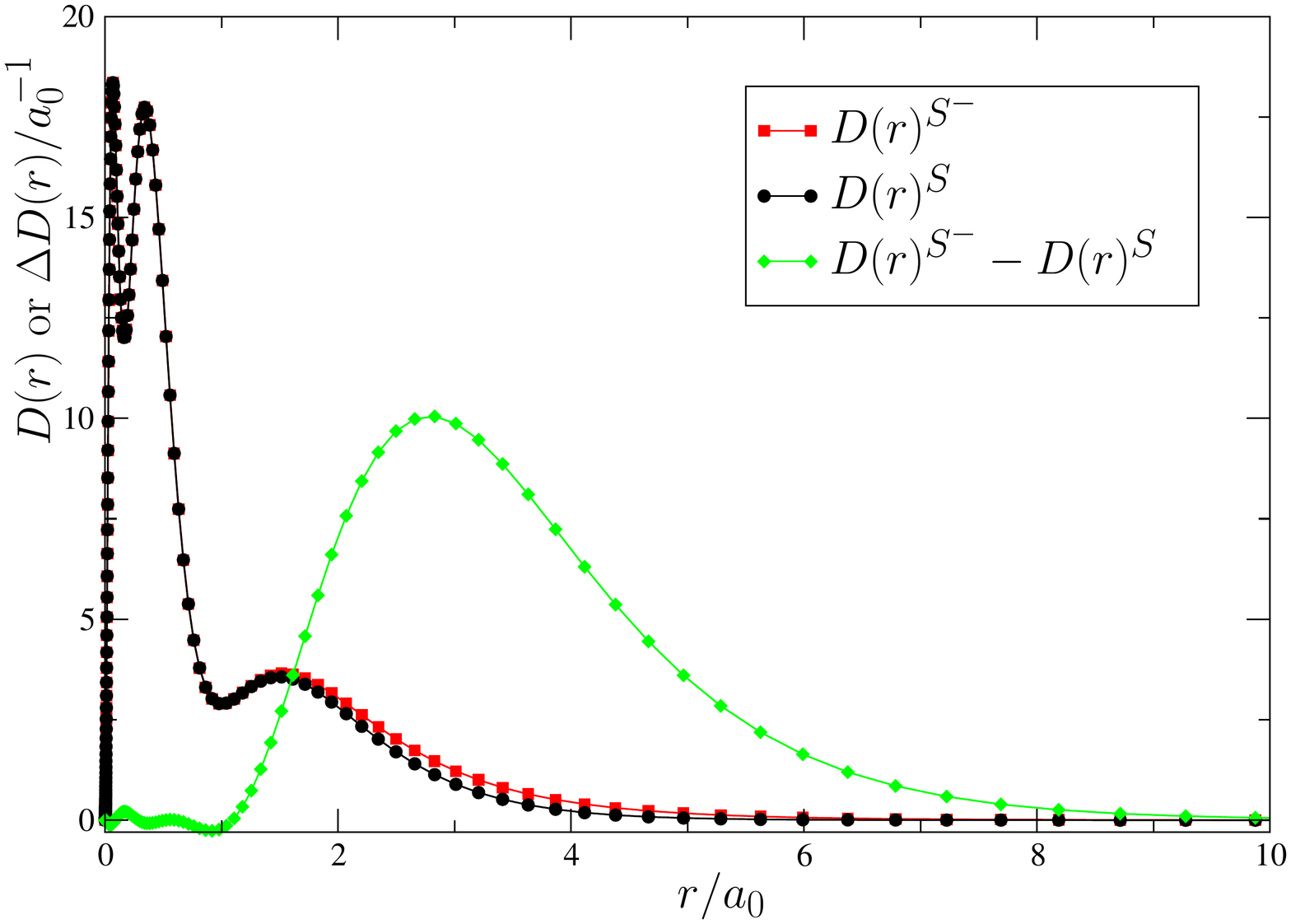}}
\caption{S and S$^-$ density functions. Density differences have been scaled by a factor 30.}
\label{fig_S}
\end{center}
\end{figure}

\section{Relativistic density functions}
\subsection{Relativistic multi-configuration wave functions}

In the relativistic scheme, the atomic wave function is, in the most general case, a combination of configuration state functions

\begin{equation} \label{Theory_Psi_MCDFeq}
\vert  \pi J M_J \rangle = \sum_{\nu} c_\nu\vert \nu \pi J M_J \rangle \; ,
\end{equation}
eigenfunction of the inversion operator $I$, the total angular momentum {\bf J}$^2$ and its projection $J_z$.  
$\nu$ denotes all the necessary information for specifying the relativistic configuration unambiguously. The CSFs are built on the one-electron Dirac four-spinor\begin{equation}
\psi_i(r) = \frac{1}{r}
\left( 
\begin{array}{c}
P_i(r) \chi_{\kappa_i}^{\mu_i} (\Omega) \\ 
i Q_i(r) \chi_{- \kappa_i}^{\mu_i} (\Omega) \end{array} \right) \, ,
\end{equation}
where $\chi_{\kappa_i}^{\mu_i} (\Omega)$ is a two-dimensional vector harmonic. It has the property that $K \psi_i(r) = \kappa \psi_i(r)$ where 
$K = \beta ( \mbox{\boldmath $ \sigma $} \cdot \bf{L} + 1).$
 The large $\{ P(r) \}$ and small $\{ Q(r) \}$ components are solutions of a set of coupled integro-differential equations~\rref{Theory_DF_integrodiff}~\cite{INDELICATO:1995uk}. The  mixing coefficients $\{c_{\nu} \}$ are obtained by diagonalizing the matrix of the no-pair Hamiltonian containing the magnetic and retardation terms \cite{Indelicato:1996by}. The two coupled variational problems are solved iteratively. For a complete discussion on relativistic atomic structure we refer to \cite{Grant:1996}. 
 
It is to be noted that the relativistic scheme rapidly becomes more complicated than the corresponding non-relativistic one.
For example, if the ground term of Carbon atom is described, in the non-relativistic one-configuration Hartree--Fock approximation, by a single CSF 
$\vert 1s^2 2s^2 2p^2 \; ^3P \rangle  $, the relativistic equivalent implies the specification of the $J$-value. For $J=0$ corresponding to the ground level of Carbon,  the following  two-configuration description becomes necessary

\begin{equation} \label{Cexpansion}
\vert `` 1s^2 2s^2 2p^2 " (J=0) \rangle = c_1 \vert 1s^2 2s^2 (2p^\ast)^2 (J=0) \rangle +  c_2 \vert 1s^2 2s^2 2p^2 (J=0) \rangle \; ,
\end{equation}

\noindent implicitly taking into account the relativistic mixing of the two $LS$-terms ($\;^1S$ and $\;^3P$) arising from the $2p^2$ configuration and belonging to the $J=0$ subspace.  $p^\ast$ and $p$ in expression~\rref{Cexpansion} correspond to the $j$-values, $j=1/2$ ($\kappa = +1$) and $j=3/2$ ($\kappa = -2$), respectively.

\subsection{Multi-configuration Dirac--Hartree--Fock equations}

For the calculations of relativistic density functions we used a multi-configuration Dirac--Fock approach (MCDF), which can be thought of as a relativistic version of the MCHF method.  The MCDF approach implemented in the MDF/GME program~\cite{DESCLAUX:1975,Des:93a} calculates approximate solutions to the Dirac equation with the effective Dirac--Breit Hamiltonian~\cite{Des:93a}

\begin{equation} \label{Theory_DFsep}
H^{DB} (\vec{r}_1, \vec{r}_2, \ldots, \vec{r}_N ) = \sum_{i=1}^{N} h^D(\vec{r}_i) + \sum_{i<j} h^B(\vec{r}_i, \vec{r}_j) \; ,
\end{equation}
with
\begin{equation}
h^D(\vec{r}_i) = c \mbox{ \boldmath $\alpha$} \cdot \mathbf{p} + c^2 (\beta - 1) V_i(r) \; 
\end{equation}
and
\begin{equation}
h^B(\vec{r}_i, \vec{r}_j) = \frac{1}{r_{ij}} - \frac{\vec{\alpha}_i \cdot \vec{\alpha}_j }{r_{ij}} \cos (\omega_{ij}r_{ij}) + \left( \alpha \cdot \vec{\nabla} \right)_i \left( \alpha \cdot \vec{\nabla} \right)_j \frac{\cos (\omega_{ij} r_{ij}) - 1}{\omega^2_{ij}r_{ij}} \; . 
\end{equation}

The total Hamiltonian contains three types of contributions: the one-electron Dirac Hamiltonion, the Coulomb repulsion and the Breit interaction. These contributions, which appear in the energy expression, give rise to radial integrals, which need to be calculated for the two component wave function. We will simply state the MCDF equations, which can be obtained by applying the variational principle to the energy expression, for variations in the expansion coefficients $c_i$ in equation~\rref{Theory_Psi_MCDFeq} and both the large and the small components of the radial wave function. The coefficients $c_i$ can be determined from the diagonalization of a Hamiltonian matrix and the radial components can be optimized by solving the coupled integro-differential equations, here given for the orbital $A$

\begin{equation} \label{Theory_DF_integrodiff}
\left[ \begin{array}{c c} \frac{d}{dr} + \frac{\kappa_A}{r} & - \frac{2}{\alpha} + \alpha V_A (r) \\ - \frac{2}{\alpha} + \alpha V_A (r) & \frac{d}{dr} + \frac{\kappa_A}{r}  \end{array}
\right] \left( \begin{array} {c} P_A(r) \\ Q_A (r) \end{array} \right) = \alpha \sum_B \varepsilon_{AB} \left( \begin{array} {c} Q_B (r) \\ -P_B (r) \end{array} \right) + \left( \begin{array} {c} X_{Q_A} (r) \\ -X_{P_A} (r) \end{array} \right) \; ,
\end{equation}
where the summation over $B$ contains only the contributions for $\kappa_A = \kappa_B$, where $V_A$ is the sum of the nuclear and the direct Coulomb potentials and where $X_{P_A}$ contains all the two electron integrals, except the instantaneous direct Coulomb repulsion. With the Lagrangian parameters $\varepsilon_{AB}$ the orthonormality constraint 
\begin{equation}
\int \left\{ P_A(r)P_B(r) + Q_A(r)Q_B(r) \right\} \; dr = \delta_{\kappa_A \kappa_B} \delta_{n_A n_B} \; ,
\end{equation}
is enforced.

\subsection{Relativistic density functions} \label{Density_DFdensity}

For the purpose of quantifying the relativistic effects on the electron density functions, which are discussed later on in this contribution, we evaluate Dirac--Fock density functions, using a point nucleus approximation. In this section we describe how density functions can be obtained.

By averaging the sublevel densities
\begin{equation}
\rho ({\bf r}) 
= \frac{1}{(2J+1)} \sum_{M_J=-J}^{+J}
\; \rho^{J M_J} ({\bf r}) \; ,
\end{equation}
the total electron density becomes spherical for any open-shell system, as found in the non-relativistic scheme (see section~\ref{secsec:3}) and can be calculated from
\begin{eqnarray} \label{relativistic_gem}
	\rho (r) & = &  \frac{1}{4 \pi} \;  
	\sum_{n \kappa} \frac{P_{n \kappa}^{2}(r) + Q_{n \kappa}^{2}(r) }{r^2} \; q_{n\kappa} \label{rel_dichtheid} \; ,
\end{eqnarray}
where $q_{n\kappa}$ is the occupation number of the relativistic subshell $(n \kappa)$. Expression~\rref{relativistic_gem} can be considered as the relativistic version of equation~\rref{av}.

\section{Analyzing atomic densities: concepts from quantum chemistry}
\subsection{The shape function} \label{Theory_shapesection}

In the context of information theory (cf. section~\ref{Chapter_Information_Theory}) the shape function, defined as the density per particle 
\begin{equation} \label{def:shape}
\sigma(\vec{r}) = \frac{\rho(\vec{r})}{N} \, ,
\end{equation}
where $N$ is the number of electrons, given by 
\begin{equation}
N = \int \rho(\vec{r}) \, d\vec{r} \, ,
\end{equation}
plays a role as carrier of information. In particular, the shape function is employed as a probability distribution which describes an atom or a molecule.

The shape function first came to the scene of quantum chemistry in 1983 with the work of Parr and Bartolotti~\cite{Parr_JPC1983}. Although the shape function had appeared before in another context, it is due to Parr and Bartolotti's work that the quantity owes its name. Just as for the density function, the shape function can be shown to determine the external potential $v(\vec{r})$ and the number of electrons $N$~\cite{Ayers:2000rc} and so completely determines the system. The Kato cusp condition~\cite{KATO:1957qq} leads to the nuclear charges and the relationship between the logarithmic derivative of the shape function and the ionization potential determines the number of electrons~~\cite{Ayers:2000rc}. On this basis a Wilson-like argument has been constructed~\cite{Geerlings:2005vf} (similar to the original DFT),  confirming the shape function ``as carrier of information"~\cite{Geerlings:2007mi}. Relationships between the shape function and concepts from conceptual DFT were established. In~\cite{ChamReact:2009gf} a slightly different perspective is given on the fundamental nature of the shape function.

\subsection{Quantum similarity} \label{Theory_Similarity}

In chemistry the similarity of molecules plays a central role. Indeed, comparable molecules, usually molecules with a similar shape, are expected to show similar chemical properties and reactivity patterns. Specifically, there chemical behavior is expected to be similar~\cite{Rouvray:1995ad}. 
The concept of functional groups is extensively used in organic chemistry~\cite{Patai:1992zi}, through which certain properties are transferable (to a certain extent) 
from one molecule to another, and the intense QSAR investigations in pharmaceutical chemistry~\cite{Bultinck:2003qw} are
illustrations of the attempts to master and exploit similarity in structure, physicochemical properties and reactivity of molecular systems.

Remarkably, the question of quantifying similarity within a quantum mechanical framework has  been addressed relatively late, in the early 1980's. The pioneering work of Carb\'{o} and co-workers~\cite{CARBO:1980,Bultinck:2005}  led to a series of quantum similarity measures (QSM) and indices (QSI). These were essentially based on the electron density distribution of the two quantum objects (in casu molecules) to be compared. The link between similarity analysis and DFT~\cite{Boon:1998rw,Boon:2001xe} built on the electron density as the basic carrier of information, and pervading 
quantum chemical literature at that time, is striking.

The last 15 years witnessed a multitude of studies on various aspects of quantum similarity of molecules (the use of different separation operators \cite{Bultinck:2005}, the replacement of the density by more appropriate reactivity oriented functions \cite{Boon:1998rw,Boon:2001xe} within the context of conceptual DFT~\cite{Geerlings:2003a}, the treatment of enantiomers~\cite{Geerlings:2005vf,Boon:2003dn,Janssens:2006tg,Janssens:2007zk}).
With the exception of two papers by Carb\'{o} and co-workers, the study of isolated atoms remained surprisingly unexplored. In the first paper~\cite{Sola:1996sf} atomic self-similarity was studied, whereas the second one~\cite{Robert:2000} contains a relatively short study on atomic and nuclear similarity, leading to the conclusion that atoms bear the highest resemblance to their neighbours in the Periodic Table.

The work discussed below is situated in the context of a mathematically rigorous theory of quantum similarity measures (QSM) and quantum similarity indices (QSI) as developed by Carb\'{o} \cite{Bultinck:2005,CARBO:1980}.
Following Carb\'{o}, we define the similarity of two atoms ($a$ and $b$) as a QSM  
\begin{equation} \label{ZABdef} 
Z_{ab}(\Omega)  =  \int  \rho_{a}(\vec{r}) 	\, \Omega (\textbf{r},\vec{r'}) \, \rho_{b}(\vec{r'}) \; d\vec{r} \, d \vec{r'} \; , 
\end{equation}
where $\Omega ({\mathbf{r}_1, \mathbf{r}_2})$ is a positive definite operator. Renormalization to
\begin{equation} \label{SIdef} 
SI_{ \Omega }  =  \frac{Z_{ab}(\Omega) } {\sqrt{Z_{aa}(\Omega)} \sqrt{Z_{bb}(\Omega) }} \;,
\end{equation}
yields a QSI $SI_{\Omega}$ with values comprised between $0$ and $1$.

The two most successful choices for the separation operator $\Omega (\textbf{r},\vec{r'})$ are the Dirac-delta $\delta(\textbf{r},\vec{r'})$ and the Coulomb repulsion $\frac{1}{|\vec{r} - \vec{r}'|}$. The first is known to reflect comparison of geometrical shape of molecules, whereas the second is said to reflect the charge concentrations~\cite{Sola:1996sf}.

\section{Analyzing atomic densities: some examples} \label{Density_QSI}
In the previous sections the calculation of the density function was discussed and a methodology for comparing them was introduced. In this section a quantitative analysis of atomic density functions is made. It seemed interesting to employ concepts from molecular similarity studies (cf.~section~\ref{Theory_Similarity}), as well developed in quantum chemistry and chemical reactivity studies. Here molecular quantum similarity measures will be applied in a straightforward fashion to investigate i) the $LS$-dependence of the electron density function in a Hartree--Fock approximation; ii)  the density functions of the atoms in their ground state, throughout the periodic table, based on the density function alone; and iii) a quantization of relativistic effects by comparing density functions from Hartree--Fock and Dirac--Hartree--Fock models.

\subsection{On the $LS$-term dependence of atomic electron density functions}

From the developments on the density function in section~\ref{secsec:3} it is clear that the $LS$-dependent restricted Hartree--Fock approximation yields $LS$-dependent Hartree--Fock equations for atoms with open sub shells. In Hartree--Fock this dependence can be traced back to the term-dependency of the coulomb interaction. In the single incomplete shell case, corresponding to the ground state configurations we are interested in for the present study, the term dependency is usually fairly small. Froese-Fischer compared mean radii of the radial functions~\cite{Fischer:1977dz}. The differences in $\langle r \rangle $  for the outer orbitals between the values obtained from a Hartree--Fock calculation on the lowest term and those for the average energy of the configuration are of the order of 1-5~\%. Although the $LS$-dependency does not show up explicitly in the direct and exchange potentials of the closed-subshell radial Hartree--Fock equations, the closed-subshell radial functions are ultimately $LS$-dependent through the coupling between the orbitals in the HF equations to be solved in the iterative procedure, but these relaxation effects turn out to be even smaller. 

As explicitly indicated through~\rref{gen} the density built from the one-electron radial functions could therefore be $LS$-dependent but this issue has not yet been investigated quantitatively. Combining the term-dependent densities 
$\rho_A = \rho_{\alpha_A L_A S_A}$ and 
$\rho_B = \rho_{\alpha_B L_B S_B}$
for the same atom in the same electronic ground state configuration, but possibly different states (and adopting the Dirac $\delta$-function for $\Omega$) for evaluating the quantum similarity measure $Z_{AB}$ of \rref{ZABdef}, the similarity matrix can be constructed according to \rref{SIdef}. Its matrix elements have been estimated for the $np^2$ configuration of Carbon $(n=2)$ and Silicon $(n=3)$~\cite{Borgoo:2004}. As expected, the deviation of the off-diagonal elements from $1$ is very small, the HF orbitals for the different terms $\; ^3P, \; ^1D$ and $\; ^1S$  being highly similar, although not identical.

\subsection{A study of the periodic table} \label{Density_QSI_noperiodicity}
 
As a first step to the recovery of the periodic patterns in Mendeleev's table, Carb\'o's quantum similarity index~\rref{SIdef} was used, with the Dirac-$\delta$ as separation operator. In this case the expression ~\rref{SIdef} reduces to an expression for shape functions~\rref{def:shape}.

For the evaluation of the QSI in expression~\rref{SIdef}, we used atomic density functions of atoms in their ground state e.g. corresponding to the lowest energy term. As elaborated in section~\ref{secsec:3} the involvement of all degenerate magnetic components allows to construct a spherical density function. For the density functions in this study, we limited ourselves to the Hartree--Fock approximation where no correlation effects are involved and the state functions are built with one CSF.

In figure \ref{CPL_noble_gas_QSI} we extract, as a 
case study from the complete atom QSI-matrix, the relevant information for the noble gases. Here the similarities were calculated 
using the Dirac delta function as separation operator. From these data 
it is clear that the similarity indices are higher, the closer the atoms 
are in the periodic table (smallest $\Delta Z$, $Z$ 
being the atomic number). The tendency noticed by Robert and Carb\'{o} in \cite{Robert:2000} 
is regained in the present study at a more sophisticated level. 
It can hence be concluded that the QSI involving $\rho(\vec{r})$ 
and evaluated with $\delta(\vec{r} - \vec{r'})$ as 
separation operator $\Omega$, does not generate periodicity.

 \begin{figure} 
 \begin{center}
      \resizebox{105mm}{!}{\includegraphics[clip,angle=0]{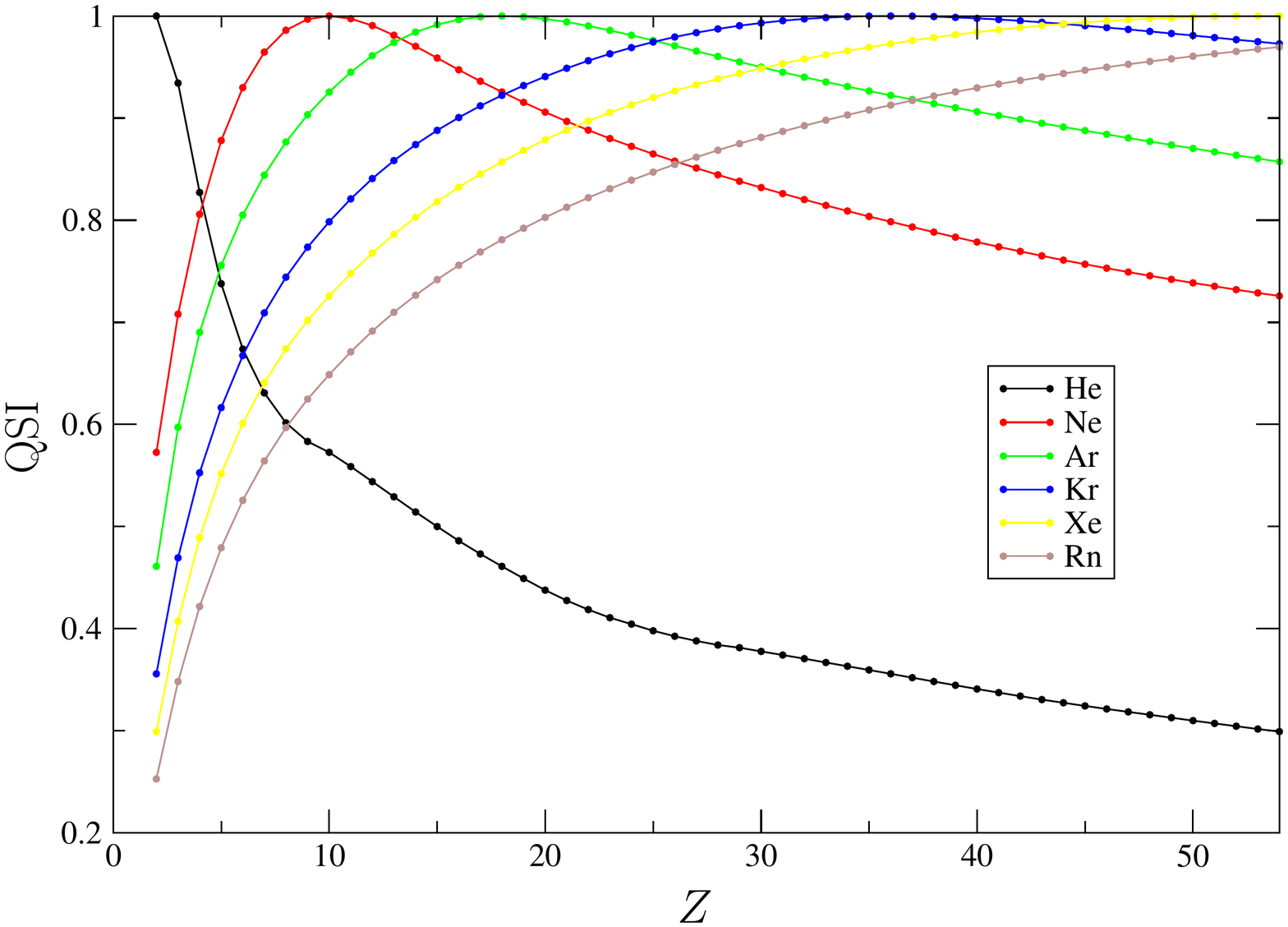}} 
 \caption{Quantum similarity indices for noble gases, using the Dirac-delta function as separation operator. 
 \label{CPL_noble_gas_QSI}}
 \end{center} 
 \end{figure}

The discussion of the work on the retrieval of periodicity is continued below in sections~\ref{CPL04} and~\ref{JCP07a}, where concepts from information theory are employed to construct a functional which quantifies the difference between two density functions in a different way.

\subsection{On the influence of relativistic effects} \label{relat}

In this section we investigate the importance of relativistic effects for the electron density functions of atoms. From the relativistic effects on total energies one can infer these effects have implications for the electron densities. 
The effect of relativity on atomic wave function has been studied in the pioneering work of Burke and Grant \cite{BURKE:1967rp} who presented graphs and tables to show the order of magnitude of corrections to the hydrogenic charge distributions for $Z=80$. The relative changes in the binding energies and expectation values of $r$ due to relativistic effects are known from the comparison of the results obtained by solving both the Schr\"odinger and Dirac equations for the same Coulomb potential. The contraction of the $ns$-orbitals is a well known example of these relativistic effects. But as pointed out by Desclaux in his ``{\it Tour historique}'' \cite{Desclaux:2002zs}, for a many-electron system, the self-consistent field effects change this simple picture quite significantly. Indeed, contrary to the single electron solution of the Dirac equation showing mainly the mass variation with velocity, a Dirac--Fock calculation includes the changes in the spatial charge distribution of the electrons induced by the self-consistent field. 

We first illustrate the difference of the radial density functions $D(r)$ defined as (see also expression~\rref{D_r})
\begin{equation}
D(r) \equiv 4 \pi r^2 \rho(\mathbf{r})  \; ,  \label{dist} 
\end{equation}
calculated in the Hartree--Fock (HF) and Dirac--Fock (DF) approximations for the ground state $6p^2 \; ^3P_0$ of Pb~I ($Z=82$) according to equations~\rref{rel_dichtheid} and~\rref{dist}, respectively. These are plotted in figure~\ref{Pbraddist}, which shows the global relativistic contraction of the shell structure.

 \begin{figure}
    \begin{center}
      \resizebox{105mm}{!}{\includegraphics[clip,angle=-90]{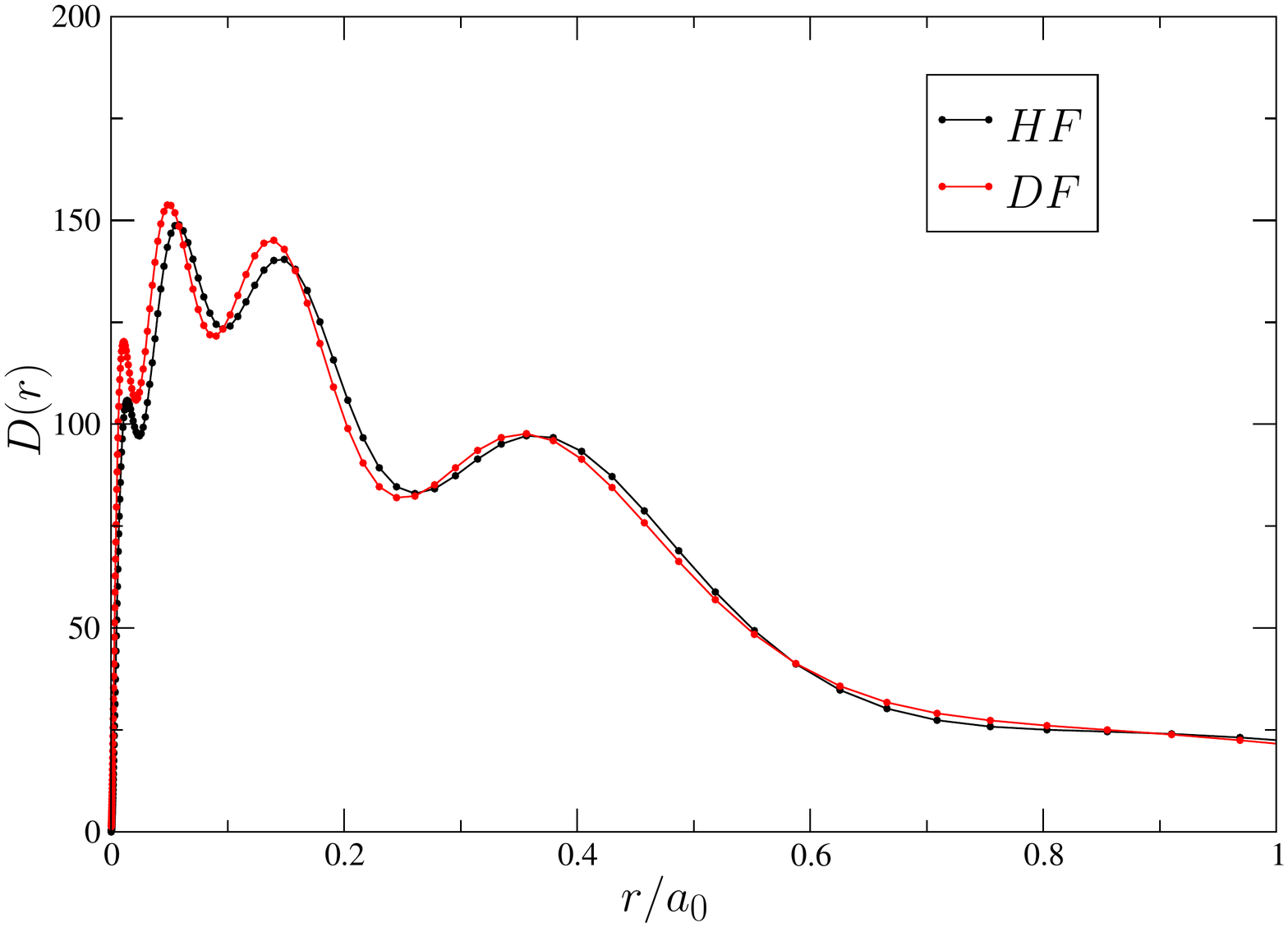}}
      \caption{DF and HF density distributions $D(r) = 4 \pi r^2 \rho(\mathbf{r}) $ for the neutral Pb atom ($Z=82$). The contraction of the first shells is clearly visible.}
      \label{Pbraddist}
    \end{center}
  \end{figure}

Employing the framework of QSI to compare non-relativistic Hartree--Fock electron density functions $\rho^{\mbox{\sc hf}}(\mathbf{r})$ with relativistic Dirac--Fock electron density functions $\rho^{\mbox{\sc df}}(\mathbf{r})$ for a given atom, the influence of relativistic effects on the total density functions of atoms can be quantified via the QSI defined as

\begin{eqnarray} 
	& Z_{\mbox{\sc hf, df}}(\delta)  =  \int  \rho^{\mbox{\sc hf}}(\mathbf{r})  \, \delta(\vec{r}-\vec{r'}) \rho^{\mbox{\sc df}}(\mathbf{r}') \; d \mathbf{r} \, d\vec{r'} \\
	& SI_{ \delta }  =  \frac{Z_ {\mbox{\sc \tiny hf, df}}(\delta) } {\sqrt{Z_ {\mbox{\tiny \sc hf, df}}(\delta) } \sqrt{Z_{\mbox{\sc \tiny hf, df}}(\delta)}} \; , \label{QSIhfvsdf}
\end{eqnarray}
where $\delta$ is the Dirac-$\delta$ operator.

In figure \ref{hfvsdf} we supply the QSI between atomic densities obtained from numerical Hartree--Fock calculation and those obtained from numerical Dirac--Fock calculations, for all atoms of the periodic table.

The results show practically no relativistic effects on the electron densities for the first periods, the influence becoming comparatively large for heavy atoms. 
To illustrate the evolution through the table the numerical results of the carbon group elements are highlighted in the graph in figure \ref{hfvsdf}. From the graph it is also noticeable that the relativistic effects rapidly gain importance for atoms heavier than Pb ($Z=86$).

 \begin{figure}
    \begin{center}
      \resizebox{120mm}{!}{\includegraphics[clip,angle=0,viewport=0 0 920 500]{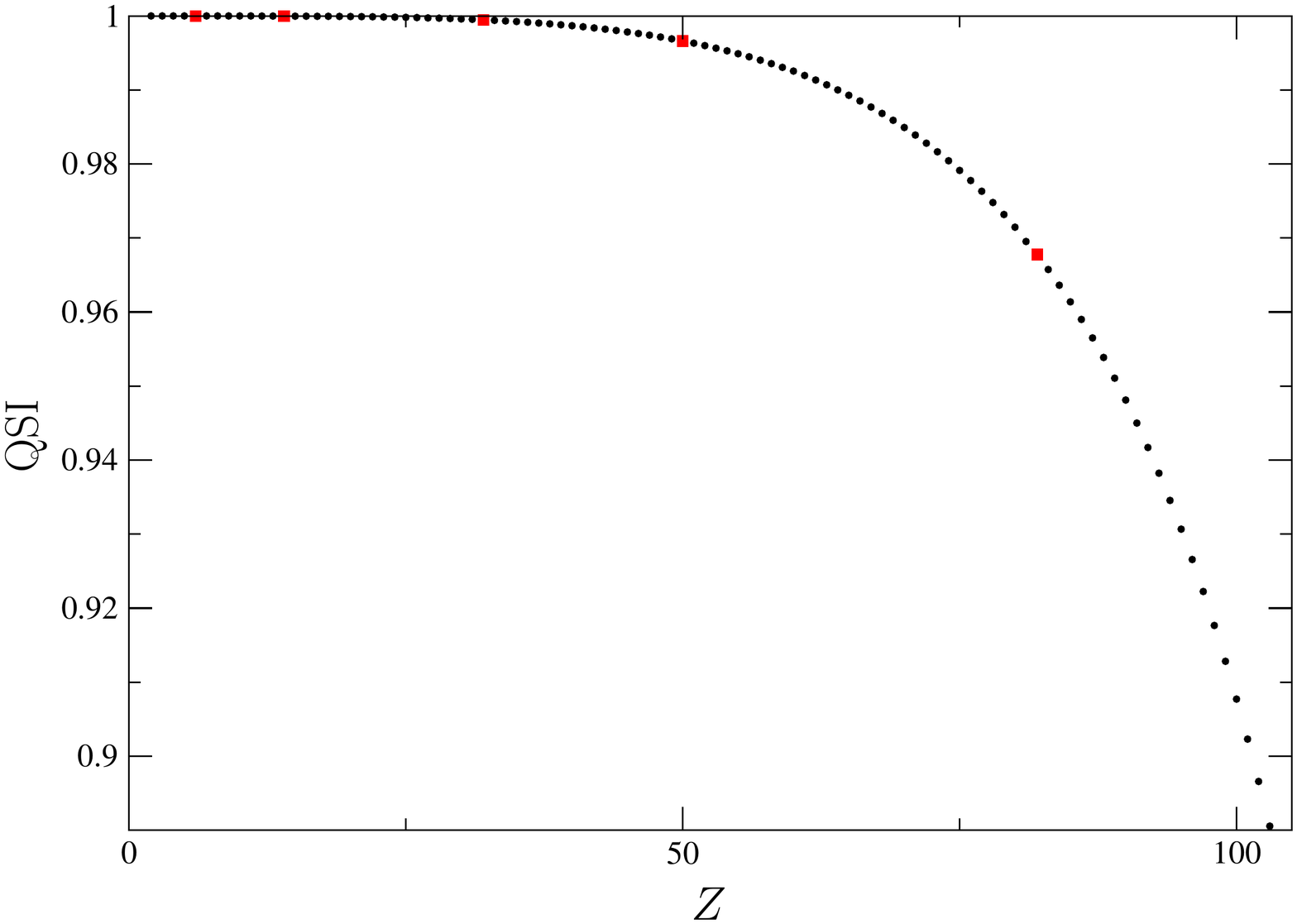}}
      \caption{Similarity of non-relativistic Hartree--Fock with relativistic Dirac--Fock atomic density functions with highlighted results for the C group atoms (C, Si, Ge, Sn, Pb).}
      \label{hfvsdf}
    \end{center}
  \end{figure}


\section{Analyzing atomic densities: concepts from Information Theory} \label{Chapter_Information_Theory}


Nowadays density functional theory (DFT) is the most widely used tool in quantum chemistry. Its relatively low computational cost and the attractive way in which chemical reactivity can be investigated made it a good alternative to traditional wave function based approaches.
DFT is based on the Hohenberg--Kohn theorems~\cite{HOHENBERG:1964}. In other words an atom's or a molecule's energy -- and in fact any other physical property -- can be determined by evaluating a density functional. 
However, the construction of some functionals corresponding to certain physical property, has proven very difficult. Moreover, to the present day no general and systematic way for constructing such functionals has been established. Although energy functionals, which are accurate enough for numerous practical purposes, have been around for some time now, the complicated rationale and the everlasting search for even more accurate energy functionals are proof of the difficulties encountered when constructing such functionals. 
In the domain of conceptual DFT, where chemical reactivity is investigated, a scheme for the construction of functionals, based on derivatives of the energy with respect to the number of electrons and/or the external potential, has proven very successful~\cite{Geerlings:2003a,Geerlings:2008PCCP}. 
Inspiration for the construction of chemically interesting functionals has also come from information theory and statistical mathematics. The functionals used for analyzing probability distributions have been successfully applied to investigate electron density functions of atoms and molecules. In this chapter we introduce those functionals and discuss several studies where they have been applied to construct chemically interesting functionals.

Shannon is generally recognized as one of the founding fathers of information theory. He defined a measure for the amount of information in a message and based on that, he developed a mathematical theory of communication. His theory of communication is concerned with the amount of information in a message rather than the information itself or the semantics. It is based on the idea that -- from the physical point of view -- the message itself is irrelevant, but its size is an objective quantity. Shannon saw his measure of information as a measure of uncertainty and referred to it as an entropy. Since Shannon's seminal publication in 1948~\cite{Shannon:1948}, information theory became a very useful quantitative theory for dealing with problems of transmission of information and his ideas found many applications in a remarkable number of scientific fields.
The fundamental character of information, as defined by Shannon, is strengthened by the work of Jaynes~\cite{Jay:1957a,Jay:1957b}, who showed that it is possible to develop a statistical mechanics on the basis of the principle of maximum entropy. 

In the literature the terms entropy and information are frequently interchanged. Arih Ben-Naim, the author of \emph{`` Farewell to Entropy: Statistical Thermodynamics Based on Information"}~\cite{Arieh-Ben-Naim:zp} insists on going one step further and motivates  ``not only to use the principle of maximum entropy in predicting the probability distribution [which is used in statistical physics], but to replace altogether the concept of entropy with the more suitable information". In his opinion ``this would replace an essentially meaningless term [entropy] with an actual objective, interpretable physical quantity [information]". 
We do not intend to participate in this discussion at this time, since the present chapter is not concerned with the development of information theory itself, but rather with an investigation of the applicability of some concepts, borrowed from information theory, in a quantum chemical context. The interested reader can find an abundance of treatments on information theory itself and its applications to statistical physics and thermodynamics in the literature. 

Information theoretical concepts found their way into chemistry during the seventies. They were introduced to investigate experimental and computed energy distributions from molecular collision experiments. The purpose of the information theoretical approach was to measure the significance of theoretical models and conversely to decide which parameters should be investigated to gain the best insight into the actual distribution. For an overview of this approach to molecular reaction dynamics, we refer to Levine's work~\cite{LEVINE:1978}. Although the investigated energy distributions have little relation with electronic wave functions and density functions, the same ideas and concepts found their way to quantum chemistry and the chemical reactivity studies which are an important study field of it. Most probably this is stimulated by the fact that atoms and molecules can be described by their density function, which is ultimately a probability distribution. 
The first applications of information theoretical concepts in quantum chemistry, can be found in the literature of the early eighties. The pioneering work of Sears, Parr and Dinur~\cite{SEARS:1980} quickly lead to more novel ideas and publications. Since then, many applications of information theoretical concepts to investigate wave functions and density functions, have been reported. 
In~\cite{Gadre:2000} Gadre gives a detailed review of the original ideas behind and the literature on ``Information Theoretical Approaches to Quantum Chemistry". To motivate our work in this field we paraphrase the author's concluding sentence: 

\begin{quote}
``Thus it is felt that the information theoretical principles will continue to serve as powerful guidelines for predictive and interpretive purposes in quantum chemistry."
\end{quote}

The initial idea in our approach was to construct a density functional, which reveals chemical and physical properties of atoms, 
since the periodicity of the Table is one of the most important and basic cornerstones of chemistry. Its recovery on the basis of the electron density alone can be considered a significant result. 
In an information theoretical context, the periodicity revealing functional can be interpreted as a quantification of the amount of information in a given atom's density function, missing from the density function of the noble gas atom which precedes it in the periodic table. The results indicate that information theory offers a method for the construction of density functionals with chemical interest and based on this we continued along the same lines and investigated if more chemically interesting information functionals could be constructed.

In the same spirit, the concept of complexity has been taken under consideration for the investigation of electron density functions. Complexity has appeared in many fields of scientific inquiry e.g.~physics, statistics, biology, computer science and economics~\cite{Catalan:2002}. At present there does not exist a general definition which quantifies complexity, however several attempts have been made. For us, one of these stands out due to its functional form and its link with information theory. 

Throughout this chapter it becomes clear that different information and complexity measures can be used to distinguish electron density functions. Their evaluation and interpretation for atomic and molecular density functions gradually gives us a better understanding of how the density function carries physical and chemical information. \emph{This exploration of the density function using information measures teaches us to read this information.}

Before going into more details about our research several concepts should be formally introduced.
For our research, which deals with applications of functional measures to atomic and molecular density functions, a brief discussion of these measures should suffice. The theoretical sections are followed by an in depth discussion of our results. 
In the concluding section we formulate general remarks and some perspectives.


\subsection{Shannon's measure: an axiomatic definition} \label{Information_Theory_Shannon}


In 1948 Shannon constructed his information measure -- also referred to as ``entropy" -- for probability distributions according to a set of characterizing axioms~\cite{Shannon:1948}. A subsequent work showed that, to obtain the desired characterization, Shannon's original four axioms should be completed with a fifth one~\cite{Mathai:1975}. Different equivalent  sets  of axioms exist which yield Shannon's information measure. The original axioms, with the necessary fifth, can be found in~\cite{ASH:1967}. Here we state the set of axioms described by Kinchin~\cite{Gadre:2000,kinchin:1957}.

For a stochastic event with a set of $n$ possible outcomes (called the event space) $\{A_1, A_2, \ldots, A_n\}$ where the associated probability distribution $P=\{ P_1, P_2, \ldots, P_n \}$ with $P_i \geq 0$ for all $i$ and $\sum_{i=1}^n P_i = 1$, the measure $S$ should satisfy:
\begin{enumerate}
\item the entropy functional $S$ is a continuous functional of $P$;
\item the entropy is maximal when $P$ is the uniform distribution i.e.~$P_i = 1/n$;
\item the entropy of independent schemes are additive i.e.~$S(P_A + P_B ) = S(P_A) + S(P_B)$ (a weaker condition for dependent schemes exists);
\item adding any number of impossible events to the event space does not change the entropy i.e.~$S(P_1, P_2, \ldots, P_n, 0, 0, \ldots, 0) = S(P_1, P_2, \ldots, P_n)$ .
\end{enumerate}
It can be proven~\cite{kinchin:1957} that these axioms suffice to uniquely characterize Shannon's entropy functional
\begin{equation} \label{entropy} 
S = - k \sum_i P_i \log P_i \; , 
\end{equation}
with $k$ a positive constant.
The sum runs over the event space i.e.~the entire probability distribution.
In physics, expression~\rref{entropy} also defines the entropy of a given macro-state, where the sum runs over all micro-states and where $P_i$ is the probability corresponding to the $i$-th micro-state.
The uniform distribution possesses the largest entropy indicating that the measure can be considered as a measure of randomness or uncertainty, or alternatively, it indicates the presence of information. 

When Shannon made the straightforward generalization for continuous probability distributions $P(x)$
\begin{equation} \label{Shannondefinition}
S[P(x)] = - k \int P(x) \log P(x) \; dx \; , 
\end{equation}
he noticed that the obtained functional depends on the choice of the coordinates. This is easily demonstrated for an arbitrary coordinate transformation $y=g(x)$, by employing  the transformation rule for the probability distribution $p(x)$
\begin{equation}
q(y) = p(x) J^{-1}
\end{equation}
and the integrandum
\begin{equation}
dy = J dx \; ,
\end{equation}
where $J$ is the Jacobian of the coordinate transformation and $J^{-1}$ its inverse.
The entropy hence becomes
\begin{equation} \label{non-invariant}
\int q(y) \log(q(y)) \; dy  = \int p(x) \log(p(x) J^{-1}) \; dx \; ,
\end{equation}
where the residual $J^{-1}$ inhibits the invariance of the entropy.
Although Shannon's definition lacks invariance and although it is not always positive, it generally performs very well. Moreover, its fundamental character is emphasized by Jaynes's maximum entropy principle, which permits the construction of statistical physics, based on the concept of information~\cite{Jay:1957a,Jay:1957b}.
In the last decade several investigations of the Shannon entropy in a quantum chemical context have been reported. 
Those relevant to our research are discussed in more detail below.



\subsection{Kullback--Leibler missing information}
Kullback--Leibler's information deficiency was introduced in 1951 as a generalization of Shannon's information entropy~\cite{Kullback:Leibler:1951}. 
For a continuous probability distribution $P(x)$, relative to the reference distribution $P_{0} (x)$, it is given by
\begin{equation} \label{KLdefinition}
\Delta S[P (x)|P_{0} (x)] = \int P (x) \log \frac {P (x)}{P_{0}(x)} \; dx \; . 
\end{equation}
As can easily be seen from expression~\rref{non-invariant}, the introduction of a reference probability distribution $P_0 (x)$ yields a measure independent of the choice of the coordinate system.
The Kullback--Leibler  functional quantifies the amount of information which discriminates $P(x)$ from $P_0(x)$. In other words, it quantifies the distinguishability of the two probability distributions. Sometimes it can be useful to see $\Delta S[P (x)|P_{0} (x)]$ as the distance in information from $P_0$ to $P$, although strictly speaking the lack of symmetry under exchange of $P(x)$ and $P_0(x)$ makes it a directed divergence.  

Kullback--Leibler's measure is an attractive quantity from a conceptual and formal point of view. It satisfies the important properties positivity, additivity, invariance, respectively:
\begin{enumerate}
\item $\Delta S[P (x)|P_{0} (x)] \geq 0$\,;
\item $\Delta S[P (x,y)|P_{0} (x,y)] = \Delta S[P (x)|P_{0} (x)] + \Delta S[P (y)|P_{0} (y)]$ for independent events i.e.~$P(x,y) = P(x) P(y)$\,;
\item $\Delta S[P (y)|P_{0} (y)] = \Delta S[P (x)|P_{0} (x)]$  if $y=f(x)$\,.
\end{enumerate}
Besides the lack of symmetry, the Kullback--Leibler functional has other formal limitations e.g.~it is not bound, nor is it always well defined.
In~\cite{Lin:1991} the lack of these properties was addressed and the Jensen--Shannon divergence was introduced as a symmetrized version of Kullback--Leibler's functional. In~\cite{Majtey:2005} the Jensen--Shannon distribution was first proposed as a measure of distinguishability of two quantum states. Chatzisavvas \emph{et al.} investigated the quantity for atomic density functions~\cite{Chatzisavvas:2005}.

For our investigation of atomic and molecular density functions, as carrier of physical and chemical information, we constructed functionals based on the definition of information measures. In sections 
\ref{CPL04} below, the research is discussed in depth. 

\section{Examples from information Theory}
\subsection{Reading chemical information from the atomic density functions} \label{CPL04}

This section contains a detailed description of our research on the recovery of the periodicity of Mendeleev's Table. The novelty in this study is that we managed to generate the chemical periodicity of Mendeleev's table in a natural way, by constructing and evaluating a density functional. As discussed before in section~\ref{Density_QSI_noperiodicity}, the comparison of atomic density functions on the basis of a quantum similarity index (using the $\delta(\vec{r}_1-\vec{r}_2) $ operator), masks the periodic patterns in Mendeleev's table. 
On the other hand, the importance of the periodicity, as one of the workhorses in chemistry, can hardly be underestimated. Due to the Hohenberg-Kohn theorems, the electron density can be considered as the basic carrier of information, although, for many properties it is unknown how to extract the relevant information from the density function. This prompted us to investigate whether the information measures, which gained a widespread attention by the quantum chemical community, could be used to help extract chemical information from atomic density functions in general and help to regain chemical periodicity in particular. 


Tempted by the interpretation of the Kullback--Leibler expression~\rref{KLdefinition} as a tool to distinguish two probability distributions, the possibility of using it to compare atomic density functions is explored. To make a physically motivated choice of the reference density $P_0(x)$ we consider the construction of Sanderson's electronegativity scale~\cite{SANDERSON:1955}, which is based on the compactness of the electron cloud. Sanderson introduced a hypothetical noble gas atom with an average density scaled by the number of electrons. This gives us the argument to use renormalized noble gas densities as reference in expression~\rref{KLdefinition}. This gives us  the quantity
\begin{equation}
	\Delta S_{A} ^{\rho} \equiv \Delta S [\rho_A(\vec{r}) | \rho_0 (\vec{r})] =  \int \rho_{A} (\vec{r}) 
	\log \frac{\sigma_{A}(\vec{r})}{\sigma_{0}(\vec{r})} \; d\vec{r} \label{KLentropy_atoms_density} \; ,
\end{equation}
where $\rho_A(\mathbf{r})$ and $\sigma_A(\mathbf{r})$ are the density and shape function of the investigated system and $\sigma_0(\mathbf{r})$ the shape function of the noble gas atom preceding atom $A$ in Mendeleev's table.
The evaluation of this expression for atoms He through Xe shows a clear periodic pattern, as can be seen in Figure~\ref{CPL04_figure2}.

Reducing the above expression to one that is based on shape functions only, leads to
\begin{equation}
	\Delta S_{A} ^{\sigma} \equiv \Delta S [\sigma_A(\vec{r}) | \sigma_0(\vec{r}) ]   =  \int \sigma_{A} (\vec{r}) 
	\log \frac{\sigma_{A}(\vec{r})}{\sigma_{0}(\vec{r})}  \; d\vec{r} \label{KLentropy_atoms_shape} \; 
\end{equation}
and its evolution is shown in Figure~\ref{CPL04_figure3}.
The periodicity is clearly present and this with the fact that the distance between 
points in a given period is decreasing gradually from first 
to fourth row is in agreement with the evolution of many chemical properties throughout the periodic table. One hereby regains one of the basic characteristics 
of the Periodic Table namely that the evolution in (many) 
properties through a given period slows down when going down 
in the Table. The decrease in slope of the four curves is a further illustration.

\begin{figure}
\begin{center}
\vspace{-2mm}
     \resizebox{105mm}{!}{\includegraphics[clip=true,angle=0]{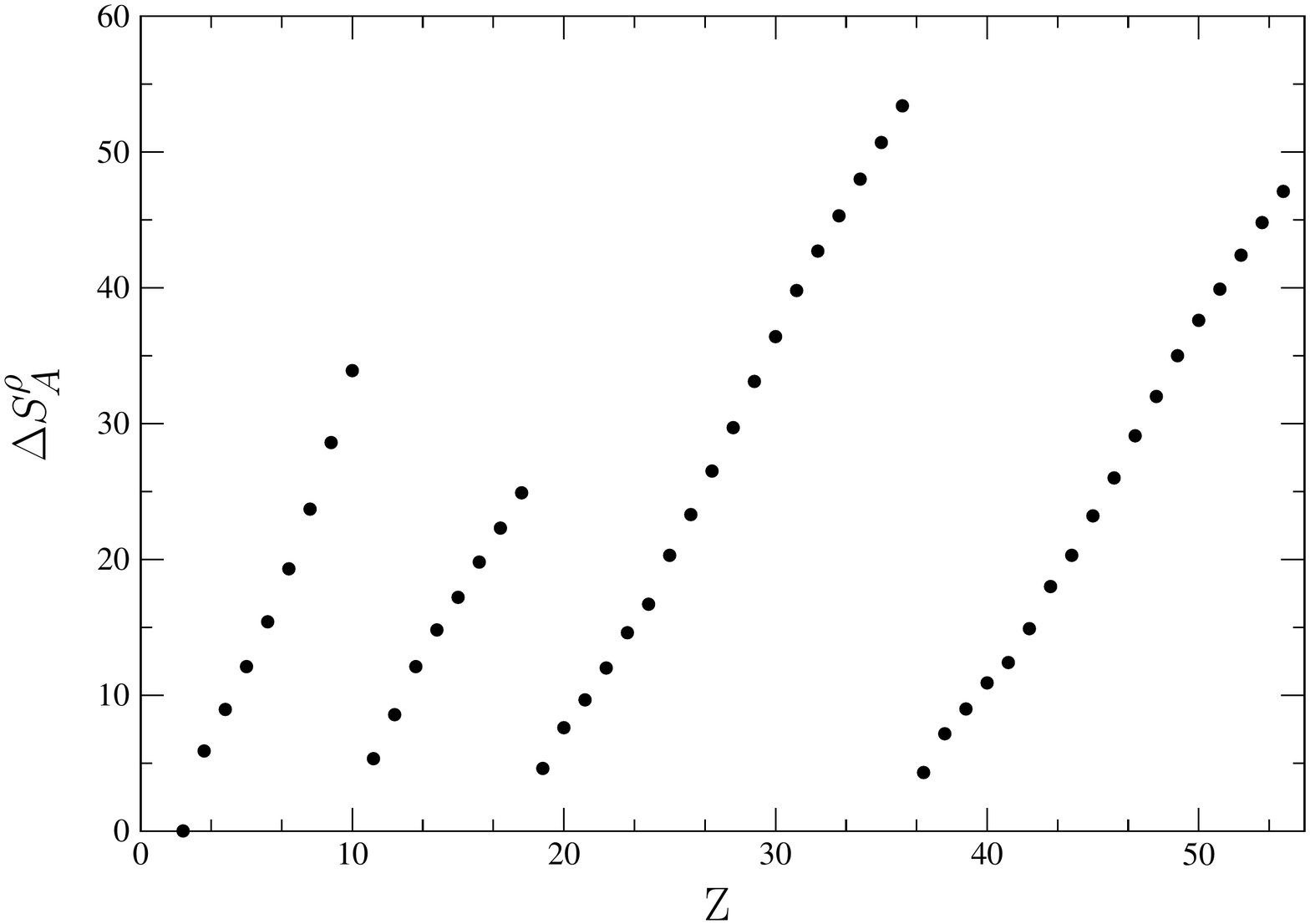}}
\caption{Kullback--Leibler information~\rref{KLentropy_atoms_density}
versus $Z$ for atomic densities with the noble gas of the previous row as reference.\label{CPL04_figure2}}
\end{center} 
 \end{figure}
 
\begin{figure}
 \begin{center}
\vspace{-4mm}
\resizebox{105mm}{!}{\includegraphics[clip,angle=-90]{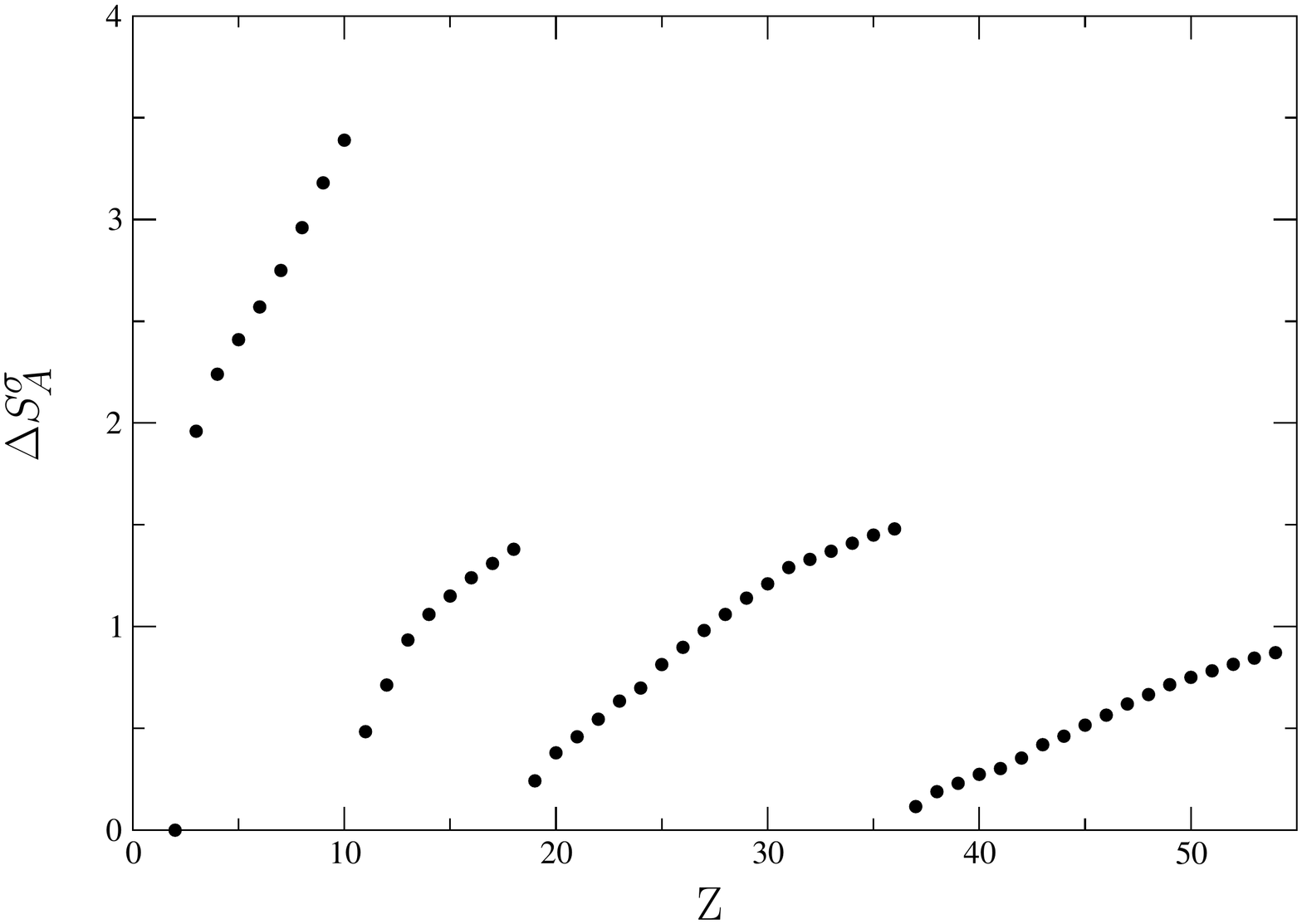}}
\caption{Kullback--Leibler information~(equation~\rref{KLentropy_atoms_shape}) 
versus $Z$ for atomic shape functions with the noble gas of the 
previous row as reference.\label{CPL04_figure3}}
\end{center}
\end{figure}

\subsection{Information theoretical QSI} \label{JCP07a}

Continuing the search for periodic patterns based on similarity measures, as introduced in section \ref{Density_QSI} and motivated by the results obtained in an information theoretical framework in section \ref{CPL04}, we will now combine the ideas from both efforts and construct an information theoretical similarity measure.


For the construction of the functional in the above section, the choice to set the reference (the prior)  density to that of a hypothetical noble gas atom, in analogy to Sanderson's electronegativity scale, was motivated and the particular choice lead to results which could be interpreted chemically.
Following these findings one can see that it would be interesting to compare the information entropy, evaluated locally as

\begin{equation}
\Delta S ^{\rho}_A(\vec{r}) \equiv \rho_{A} (\vec{r}) \log \frac{\rho_{A}(\vec{r})}{\frac{N_{A}}{N_{0}}\rho_{0}(\vec{r})} \; ,
\end{equation}

\noindent for two atoms by use of a QSM, which can be constructed straightforwardly, by considering the overlap integral 
(with Dirac-$\delta$ as separation operator) of the local information entropies of two atoms $A$ and $B$ 

\begin{equation}   
	Z_{AB}(\delta)  =  \int  \rho_{A} (\vec{r}) \log \frac{\rho_{A}(\vec{r})}{\frac{N_{A}}{N_{0}}\rho_{0}(\vec{r})} \rho_{B} (\vec{r}) \log \frac{\rho_{B}(\vec{r})}{\frac{N_{B}}{N_{0'}} \rho_{0'}(\vec{r})} \; d \vec{r}  \; .
\end{equation}

A QSI can be defined by normalizing the QSM as before, via expression~\rref{SIdef}. 
The QSM and the normalized QSI give a quantitative way of studying the resemblance in the information carried by the valence electrons of two atoms. 
The obtained QSI trivially simplifies to a shape based expression
\begin{equation} \label{entropyQSI}
SI_{(\delta)}=\frac{\int \Delta S_A ^\sigma (\vec{r}) \Delta S_B ^\sigma (\vec{r}) d \vec{r}} {\sqrt{ \int \Delta S_A ^\sigma (\vec{r}) \Delta S_A ^\sigma (\vec{r}) d \vec{r}} \sqrt{\int \Delta S_B^\sigma(\vec{r}) \Delta S_B^\sigma(\vec{r})d \vec{r}}} \; .
\end{equation}


To illustrate the results we select the QSI~\rref{entropyQSI} with the top atoms of each column as prior. Formulated in terms of Kullback--Leibler information discrimination the following is evaluated.
For instance, when we want to investigate the distance of the atoms Al, Si, S and Cl from the N-column (group Va), we consider the information theory based QSI in expression~\rref{entropyQSI}, where the reference densities $\rho_0$ and $\rho_{0'}$ are set to $\rho_N$, $\rho_A$ to $\rho_{Al}$, $\rho_{Si}$,  $\rho_{P}$, etc. respectively and $\rho_B$ to $\rho_{P}$, i.e.~we compare the information contained in the shape function of N to determine that of P, with its information on the shape function of Al, Si, S, Cl. Due to the construction a $1.$ is yielded for the element P and the other values for the elements to the left and to the right of the N-column decrease, as shown in figure~\ref{plotcol}. This pattern is followed for the periods $3$ up to $6$, taking As, Sb and Bi as reference, with decreasing difference along a given period in accordance with the results above. Note that the difference from $1.$ remains small, due to the effect of the renormalization used to obtain the QSI.
\vspace{-10mm}
\begin{figure}
   \begin{center}
           \resizebox{110mm}{!}{\includegraphics[clip=true,angle=0]{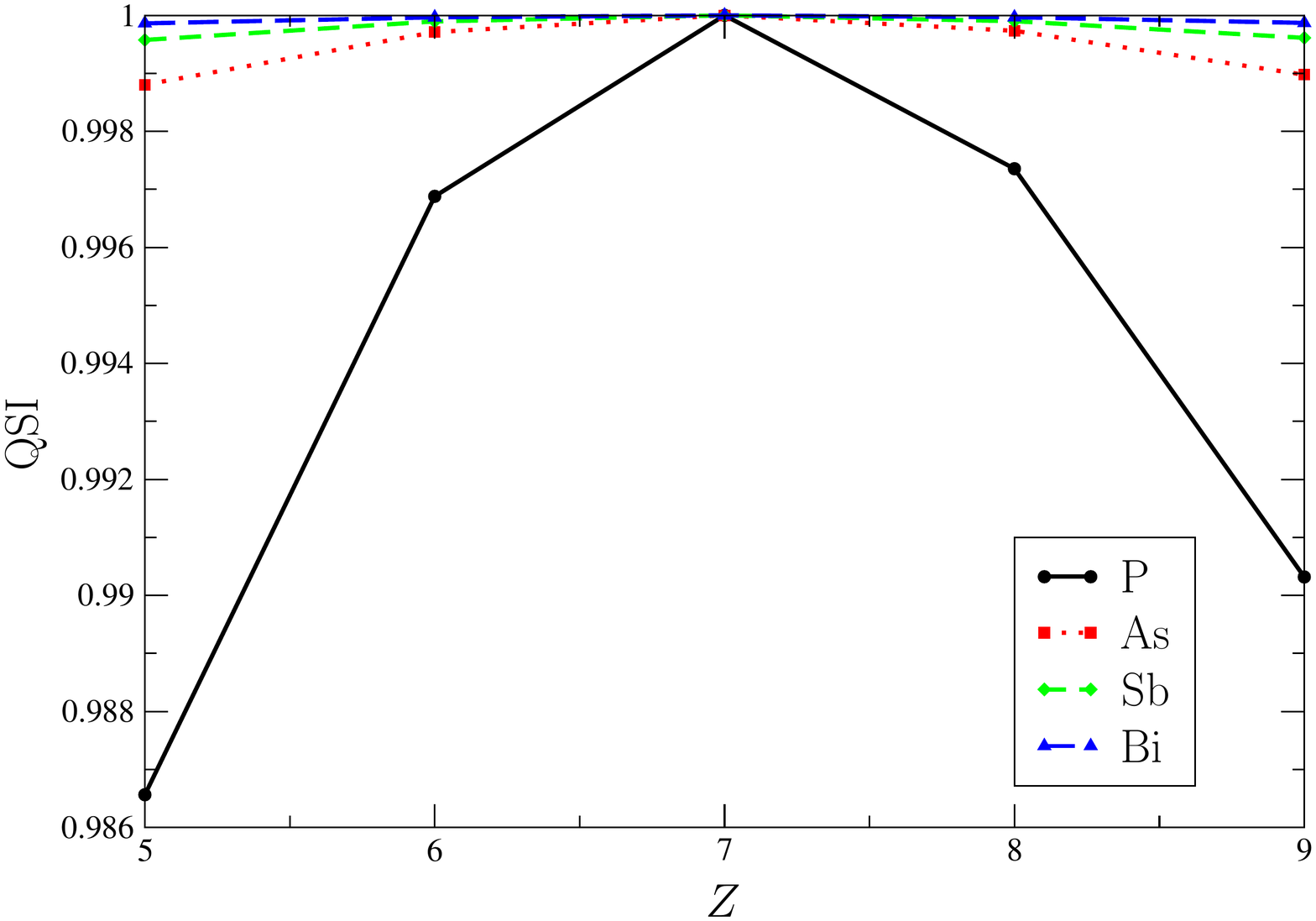}}
     \caption{Results of the information theory based QSI with the atom on top of the column as prior. The symbol in the legend indicates the period of the investigated atom and the nuclear charge $Z$-axis indicates the column of the investigated atom. (For example Ga can be found as a square $Z=5$). }
     \label{plotcol}
   \end{center}
 \end{figure}

%


\section{General Conclusion}

Results on the investigation of atomic density functions are reviewed. First, ways for calculating the density of atoms in a well-defined state are discussed, with particular attention for the spherical symmetry. It follows that the density function of an arbitrary open shell atom is not a priori spherically symmetric. A workable definition for density functions within the multi-configuration Hartree--Fock framework is established. By evaluating the obtained definition, particular influences on the density function are illustrated. A brief overview of the calculation of density functions within the relativistic Dirac--Hartree--Fock scheme is given as well.

After discussing the definition of atomic density functions, quantum similarity measures are introduced and three case studies illustrate that specific influences on the density function of electron correlation and relativity can be quantified in this way. Although no periodic patterns were found in Mendeleev's table, the methodology is particularly successful for quantifying the influence of relativistic effects on the density function.

In the final part the application of concepts from information theory is reviewed. After covering the necessary theoretical background a particular form of the Kullback--Liebler information measure is adopted and employed to define a functional for the investigation of density functions throughout Mendeleev's Table. The evaluation of the constructed functional reveals clear periodic patterns, which are even further improved when the shape function is employed instead of the density functions. These results clearly demonstrate that it is possible to retrieve chemically interesting information from the density function. Moreover the results indicate that the shape function further simplifies the density function without loosing essential information. The latter point of view is extensively treated in~\cite{C0CP01046D}, where the authors elaborately discuss ``information carriers'' such as the wave function, the reduced density matrix, the electron density function and the shape function.

 \bibliographystyle{pccp}
 \bibliography{bibliography}

\end{document}